\documentclass{tex/src/journal/aa}

\newcommand{\projecttitle}{NASIM: Revealing the low surface brightness Universe from legacy VISTA data}
\newcommand{\projectversion}{9d9968e}
\newcommand{\projectgitrepo}{https://gitlab.com/nasim-projects/pipeline}
\newcommand{\projectgitbranch}{paper}
\newcommand{\projectcopyrightowner}{Elham Saremi <saremy.elham@gmail.com>}
\newcommand{\maneagedate}{22 May 2023}
\newcommand{\maneageversion}{8161194}
\newcommand{\machinearchitecture}{x86\_64}
\newcommand{\machinebyteorder}{Little Endian}

\ifdefined\highlightnew

\else

\fi

\ifdefined\highlightnotes
\newcommand{\tonote}[1]{\textcolor{red!60!black}{[#1]}}
\else
\newcommand{\tonote}[1]{{}}
\fi

\usepackage{graphicx}

\usepackage{fixltx2e}

\usepackage{xcolor}
\color{black} 
\definecolor{DarkBlue}{RGB}{0,0,90}

\usepackage{setspace, caption}
\usepackage[
  colorlinks,
  urlcolor=blue,
  citecolor=blue,
  linkcolor=blue,
  linktocpage]{hyperref}

\hypersetup{
    pdftitle={\projecttitle},
    pdfauthor={\projectcopyrightowner},
    pdfsubject={\projectgitrepo{} (commit \projectversion)},
    pdfkeywords={Reproducible research, Maneage, ADD YOUR OWN}
}

\usepackage{tikz}
\usetikzlibrary{external}
\tikzsetexternalprefix{tex/tikz/}
\usepackage{xcolor}
\usepackage{mathptmx}
\usepackage[utf8]{inputenc}

\usepackage{pgfplots}
\pgfplotsset{compat=newest}
\usepgfplotslibrary{groupplots}
\pgfplotsset{
  axis line style={thick},
  tick style={semithick},
  tick label style = {font=\footnotesize},
  every axis label = {font=\footnotesize},
  legend style = {font=\footnotesize},
  label style = {font=\footnotesize}
  }

\usepackage{txfonts}

\title{\projecttitle}
\subtitle{}

 \titlerunning{NASIM: Advanced LSB reduction for VISTA}

\begin{document}

\author{Elham Saremi\inst{1,2,3},
          Ignacio Trujillo\inst{1,2},
          Mohammad Akhlaghi\inst{4},
          Zohreh Ghaffari\inst{5,6},
          Johan H. Knapen\inst{1,2},
          Manda Banerji\inst{3},
          Helmut Dannerbauer\inst{1,2},
          \and
          Sébastien Comerón\inst{2,1}
          }

   \institute{Instituto de Astrofísica de Canarias, C/ Vía Láctea
              s/n, 38205 La Laguna, Tenerife, Spain; \email{e.saremi@soton.ac.uk}
         \and
             Departamento de Astrofísica, Universidad de La Laguna, 38205 La Laguna, Tenerife, Spain
         \and
             School of Physics \& Astronomy, University of Southampton, Highfield Campus, Southampton SO17 1BJ, UK
         \and
             Centro de Estudios de Física del Cosmos de Aragón (CEFCA), Unidad Asociada al CSIC, Plaza San Juan 1, 44001 Teruel, Spain
         \and
             INAF-Osservatorio Astronomico di Trieste, Via G. B. Tiepolo 11, 34143 Trieste, Italy
        \and
             IFPU, Institute for Fundamental Physics of the Universe, via Beirut 2, 34151 Trieste, Italy
             }

    \authorrunning{E. Saremi et al.}
   \date{}

\abstract
  {Near-infrared imaging is a powerful technique in observational astronomy, but the bright background—primarily from the Earth's atmosphere—makes the detection of faint features particularly challenging. To recover low surface brightness (LSB) structures in such data, we present NASIM (Near-infrared Automated low Surface brightness reduction In Maneage), a fully automated and reproducible data reduction pipeline optimised for VISTA/VIRCAM observations. NASIM builds on advanced techniques from the GNU Astronomy Utilities (Gnuastro) to effectively remove large-scale instrumental artefacts and detector patterns while preserving faint, diffuse emission crucial for LSB science.

  As a key science application, we focus on the deep $K_s$-band observations of the Euclid Deep Field South (KEDFS), one of the deepest VISTA/VIRCAM datasets and a high-priority field for synergy with current and future facilities, including Euclid, JWST, LSST, Roman, Spitzer, and ALMA. Situated near the South Ecliptic Pole, KEDFS offers favourable conditions for deep extragalactic observations due to minimal Galactic foreground contamination. With VIRCAM no longer operational, it now stands as a unique and irreplaceable legacy dataset for near-infrared astronomy.

  We release selected tiles from the KEDFS survey and highlight representative science cases—including galaxy outskirts, LSB galaxies, and intracluster light—that demonstrate NASIM’s ability to recover diffuse structures. Our reduction reaches a surface brightness limit of $\sim$27.7\,mag\,arcsec$^{-2}$ ($3\sigma$ over 100 arcsec$^2$) in the $K_s$ band, approximately 67 times deeper than 2MASS and 11 times deeper than VISTA Hemisphere Survey (VHS). A direct comparison with conventional VISTA data reduction pipelines demonstrates the advantages of NASIM in preserving diffuse emission without compromising compact source detection. All quantitative results presented in this paper are fully reproducible with Maneage (commit \texttt{\projectversion}).}


\maketitle


\begingroup
\addtolength{\topskip}{0.5cm}
\begin{table*}
\centering
\begin{tabular}{lllcll}
  \hline
  \hline
Survey & Area       & Filters & Depth ($K_s$)        & Science Goals  & Reference \\
       & (deg$^2$)  &         & 5$\sigma$; AB-mag  &                &           \\
\hline
VHS     & 20,000 & $J, K_s$    & $\sim$20& Uniform southern hemisphere NIR imaging & \cite{McMahon13} \\
VIKING  & 1500   & $Z, Y, J, H, K_s$ & $\sim$21.2 & Samples galaxies at z $\gtrsim$ 1  & \cite{Edge13} \\
VISIONS & 650    & $J, H, K_s$ & $\sim$19.5 & Five nearby star-forming molecular clouds&\cite{Meingast23} \\
VVV     & 520    & $Z, Y, J, H, K_s$ & $\sim$20.0 & Variable stars in the Milky Way  & \cite{Minniti10} \\
SHARKS  & 300    & $K_s$       & $\sim$22.7 & The most massive structures in the Universe  & \cite{Dannerbauer22} \\
VMC     & 184    & $Y, J, K_s$ & $\sim$21.1 & Stellar populations in Magellanic Clouds & \cite{Cioni11} \\
GCAV    & 38     & $Y, J, K_s$ & $\sim$22.4  & Observing 20 Clusters of Galaxies & \href{https://archive.eso.org/cms/eso-archive-news/third-data-release-from-the-galaxy-clusters-at-vircam-gcav-eso-vista-public-survey.html}{GCAV DR3; 2023} \\
KEDFS   & 20     & $K_s$             & $\sim$23.5 & Euclid Deep follow-up in the $K_s$ band   & \href{https://archive.eso.org/wdb/wdb/eso/abstract/query?progid=106.21LU.001}{KEDFS; Abstract} \\
VIDEO   & 12     & $Z, Y, J, H, K_s$ & $\sim$23.5 & Galaxy evolution to z $\sim$4; massive galaxies &\cite{Jarvis13} \\
VEILS   & 9      & $J, K_s$       & $\sim$22.5 & Reionisation studies and massive galaxies & \cite{Honig17} \\
UltraVISTA & 2.2 & $Y, J, H, K_s$ & $\sim$24.0 & Ultra-deep NIR survey in COSMOS field  & \cite{McCracken12} \\
\hline
\end{tabular}
\vspace{0.1cm}
\caption{\vspace{0.25cm}Summary of selected VISTA surveys, sorted by area.}
\label{tab:vista}
\end{table*}
\endgroup

\section{Introduction}

To date, the low surface brightness (LSB) Universe (typically defined as regions with $\mu_K > 24$ mag arcsec$^{-2}$) has not been thoroughly explored in the near-infrared (NIR) from the ground, primarily due to the significant challenges posed by atmospheric emission.
A strong thermal background produces a typical sky brightness of $\mu_K \sim 14.85$ mag arcsec$^{-2}$ \citep{Cuby00}, nearly 1000 times brighter than the optical sky \citep{Trinh13}.
This overwhelming background introduces substantial noise that hinders the detection of faint, extended structures \citep{Maihara93}.
These limitations are particularly severe for wide-field surveys, where the sky varies across the field and over time, complicating both calibration and sky subtraction \citep{Borlaff19}.
As a result, astronomers have traditionally avoided targeting diffuse NIR features (such as galaxy outskirts or LSB dwarfs) from the ground.
Nonetheless, ground-based facilities offer distinct advantages, including larger mirror apertures, more versatile instrumentation, and easier access for maintenance and upgrades, which make them essential for advancing deep NIR observations despite atmospheric challenges.

The NIR is a powerful window into galaxy evolution, particularly for tracing the old stellar populations that dominate galaxy mass \citep{Zibetti09,Conroy13}.
Within this regime, the $K$ band (centred at $2.2\,\mu\mathrm{m}$) is especially valuable, offering a stable and direct tracer of stellar mass across a wide range of galaxy types and redshifts \citep[e.g.][]{Grosbol04,Kendall08}.
Unlike optical wavelengths, which are more sensitive to recent star formation and dust attenuation, $K$-band light is dominated by emission from older, low-mass stars and is relatively unaffected by star formation history, metallicity, or extinction \citep{Kochanek01, Bell03, Drory04, McGaugh14}.
This makes $K$-band selection highly effective for identifying the most massive galaxies up to $z \sim 2$, where rest-frame $K$-band light continues to trace stellar mass reliably \citep{Caputi05}.
As such, the $K$-band surveys prove especially powerful for constraining the formation epochs and assembly histories of early massive galaxies.

The $K$ band is also instrumental for probing galaxy environments: galaxies selected in this band tend to be more strongly clustered than their optically selected counterparts, reflecting a bias towards massive systems residing in dense regions of the cosmic web \citep{Furusawa11}.
Environmental studies using marked correlation functions confirm that $K$-band luminosity of galaxies exhibits a measurable dependence on local galaxy density, reinforcing its value as a mass-sensitive environmental probe \citep{Deng20, Sureshkumar21}.

In this decade of transformative astrophysical surveys, Euclid is poised to deliver unprecedented NIR imaging in the $Y$, $J$, and $H$ bands from space, offering high-resolution, wide-area coverage of the extragalactic sky \citep{Euclid25}.
These data are further complemented by mid-infrared observations from the Spitzer Space Telescope (Spitzer), enabling powerful constraints on galaxy evolution over cosmic time \citep{Euclid22}.
However, despite this extensive coverage, a key gap remains in the $K$ band, crucial for filling the spectral continuum and for connecting observations across redshift and stellar population regimes.

Fortunately, this gap is being effectively bridged by ground-based facilities: UKIRT offers valuable $K$-band data in the northern hemisphere \citep{Lawrence07, Hambly08, Schneider25}, and the VISTA telescope, the largest ground-based NIR survey instrument, delivered comprehensive southern sky coverage \citep{Emerson06}.
Through its deep and wide-area programmes, VISTA/VIRCAM (now decommissioned) plays a pivotal role in completing the multi-wavelength coverage of current and future surveys, with particular strength in the $K_s$ band.
In addition to this critical spectral contribution, VISTA/VIRCAM supports a diverse portfolio of public legacy surveys, each designed to address complementary aspects of cosmology and galaxy evolution.
These surveys span a wide range of depths and sky areas, from shallow all-sky mapping to narrow ultra-deep fields, and collectively target a broad set of science goals (see Section \ref{sec:data} and Table \ref{tab:vista}).

To fully capitalise on VISTA’s rich survey data, sophisticated processing pipelines are essential.
The VISTA Data Flow System (VDFS; \citealt{Emerson04}), developed and maintained by the Cambridge Astronomy Survey Unit (CASU) and the Wide Field Astronomy Unit (WFAU), provides the core infrastructure for reducing and calibrating VISTA images \citep{Irwin04, Cross12}.
CASU handles the initial data reduction, including flat-fielding, background subtraction, astrometric and photometric calibration, while WFAU oversees catalogue generation and database curation \citep{Cross12}.
These pipelines are optimised for point sources and compact galaxies, providing robust measurements for a broad range of scientific applications.

However, traditional processing pipelines often struggle to preserve diffuse LSB structures.
Sky subtraction techniques and source detection algorithms, optimised for compact sources, can inadvertently suppress or remove extended emission.
This limitation is particularly critical for studies focussing on galaxy outskirts, stellar haloes, tidal streams, and intracluster light (ICL); components that are essential for understanding hierarchical galaxy assembly and the physics of baryonic matter in low-density environments \citep[e.g.][]{MontesTrujillo14, FliriTrujillo16, Borlaff19, Roman21}.

To address these limitations, we have developed NASIM (Near-infrared Automated low Surface brightness reduction In Maneage), a dedicated data reduction pipeline optimised for LSB science with VISTA/VIRCAM.
NASIM effectively removes large-scale instrumental signatures and flat-field patterns inherent to VIRCAM’s complex detector layout, while preserving the faint, extended features essential for LSB science.
It provides a complementary, LSB-optimised alternative to the standard VDFS outputs, enabling high-fidelity exploration of diffuse structures in the NIR.

This paper is structured as follows.
In Section \ref{sec:data}, we introduce several VISTA surveys and, in particular, describe the specific dataset used in this study.
Section \ref{sec:pipeline} presents the NASIM pipeline, detailing its architecture, data reduction steps, and reproducibility framework.
In Section \ref{sec:showcase}, we showcase a selection of science cases enabled by NASIM, highlighting its effectiveness in revealing faint, extended structures in the NIR.
Section \ref{sec:discussion} provides a comparative assessment between NASIM and standard reduction pipelines, demonstrating its advantages for LSB science.
Finally, Section \ref{sec:summary} presents a brief conclusion.
All magnitudes in this paper are expressed in the AB photometric system.

\section{Data} \label{sec:data}

The VISTA telescope, equipped with the VIRCAM instrument \citep{Sutherland15}, has enabled a suite of wide-field NIR public surveys across the southern sky.
These include large-area surveys such as the VISTA Hemisphere Survey (VHS), medium-deep campaigns like the VISTA Kilo-degree Infrared Galaxy (VIKING) survey, deeper extragalactic programmes such as VIDEO and UltraVISTA, and various other legacy surveys with diverse scientific goals.
Each survey is designed with distinct scientific goals and spans a range of depths and sky areas, as summarised in Table \ref{tab:vista}.
Despite their diversity, a common challenge across these surveys is the recovery of extended LSB emission, a science regime where traditional pipelines often prove inadequate or lack the sensitivity required for reliable detection.
As detailed in Section \ref{sec:pipeline}, the NASIM pipeline is developed to address this limitation and is broadly compatible with VIRCAM data from any of these surveys.

In this paper, we apply NASIM to a new VISTA Large Programme for the first time: the KEDFS survey (Programme ID: 106.21LU.001; PI: Nonino, M.), a deep $K_s$-band campaign over the Euclid Deep Field South (EDFS).
KEDFS spans approximately 20 deg$^2$, centred at R.A.\,(J2000) = 04$^{\mathrm{h}}$\,04$^{\mathrm{m}}$\,57$^{\mathrm{s}}$.84 and Dec\,(J2000) = $-48^\circ$\,25$'$\,22$''$.8, near the south ecliptic pole.
This region is a strategically chosen deep field with exceptional current or future multi-wavelength coverage: high-resolution $Y$, $J$, and $H$ imaging from Euclid \citep{Euclid23}, deep optical data from the Vera C. Rubin Observatory \citep{Gris24}, and mid-infrared imaging from Spitzer \citep{Euclid22}.
Additionally, the field lies within the continuous viewing zones of both Roman (formerly WFIRST) and JWST, further enhancing its long-term scientific value.
The EDFS’s location near the south ecliptic pole also makes it uniquely advantageous, offering low Galactic extinction, low stellar density, and an absence of bright foreground stars, all favourable conditions for extragalactic science.

The KEDFS survey is designed to bridge the critical wavelength gap between Euclid and Spitzer by providing $K_s$-band data with VIRCAM.
It consists of 15 tiles, each constructed from a six-pointing ``tile'' sequence that mosaics the sixteen non-contiguous VIRCAM detectors (``pawprints'') into contiguous 1.5$\times$1 deg$^2$ fields.
In total, 207 hours of integration were collected between December 2020 and November 2022.
These observations offer a unique combination of depth and area in the NIR, enabling robust studies of ultra-massive passive galaxies, high-redshift galaxy clusters, and faint diffuse features.

With VIRCAM no longer in operation, KEDFS stands as a unique and irreplaceable legacy dataset.
Its depth, area, and spectral placement make it an ideal test case for the NASIM pipeline, which aims to unlock the LSB science potential in this and other VISTA surveys.

\section{NASIM} \label{sec:pipeline}

We developed a specialised pipeline tailored to study LSB structures, named NASIM (Near-infrared Automated low Surface brightness reduction In Maneage).
NASIM is built on the fully reproducible workflow framework Maneage \citep{maneage}, which is implemented in GNU Make.
Data lineage is tracked through Make's dependency system, which avoids repeating completed steps, enables parallel execution, and ensures full traceability of all processing stages.
Maneage autonomously downloads and configures all required software and libraries for NASIM within a self-contained environment, independent of the host operating system.
Additionally, the entire project — including the software environment, analysis pipeline, and and text of this paper — is version-controlled with Git, preserving the full history of all changes.
This methodology ensures long-term transparency and reproducibility.

\begin{figure*}[t]
  \begin{center}
  \ifdefined\makepdf%
    \tikzsetnextfilename{figure-structure-pipeline}%
    \input{tex/src/figure-structure-pipeline.tex}%
  \else
    \includegraphics[width=0.95\linewidth]{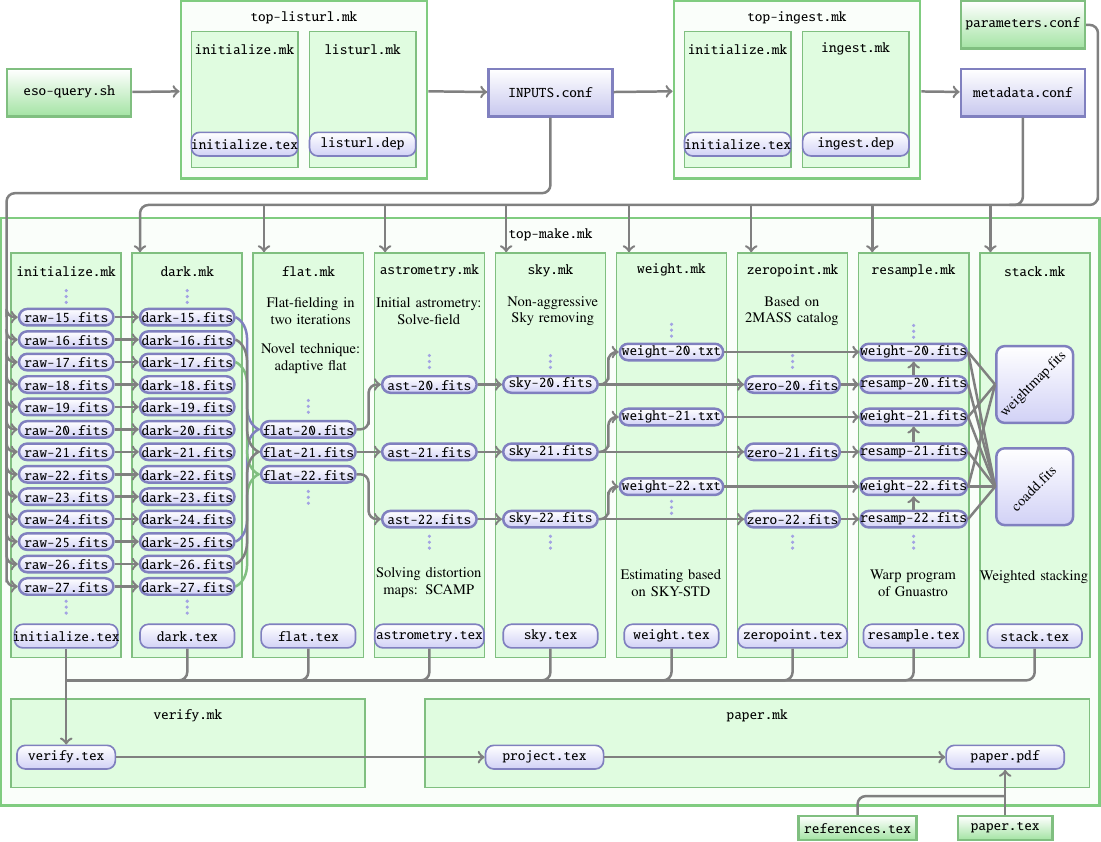}
  \fi

  \end{center}
  \vspace{-3mm}
  \caption{\label{fig:structure}
    Structural diagram of the NASIM pipeline. Each coloured box represents a file, with arrows indicating the flow of inputs to outputs. Green boxes correspond to plain-text source files that are version-controlled within the project directory, while blue boxes represent output products generated during the build process. The Makefiles (\texttt{*.mk}) define the targets and dependencies that govern the automated execution of each pipeline stage.
   }
\end{figure*}

The entire process in the NASIM pipeline can be categorised into three top Makefiles.
The initial two preparation Makefiles pertain to data download and classification, while the primary focus on data release is encapsulated within the main top Makefile.
A schematic representation of NASIM is shown in the Figure \ref{fig:structure}.
Each coloured box represents a file in the project, with arrows indicating the flow from input to output files.
Green boxes represent plain-text files that are part of the version control system and are located within the project's source directory.
These files include essential Makefiles, specific configuration files, and the actual text of the paper, all of which are manually crafted.
Blue boxes denote output files located in the build directory, which are specifically defined as targets within the Makefiles.
Consequently, these files are generated automatically.

\subsection{Data preparation phase}

The ESO search engine provides a query script based on the programme ID and tile name.
The first preparation Makefile, \texttt{top-listurl.mk} (top-left box in Figure \ref{fig:structure}), generates a file \texttt{INPUTS.conf}.
This file which is under version control captures the URLs of raw images and their corresponding SHA256 checksums, using the ESO query script as a reference.
The SHA256 checksum is crucial for automated verification, ensuring that the file remains unchanged across different project runs (both locally and through URLs), thereby maintaining the integrity of the reproducible pipeline.

In the second phase of preparation, \texttt{top-ingest.mk} (top-right box in Figure \ref{fig:structure}), the data download process is executed.
This phase focusses on retrieving raw science images, dark images, and bad pixel masks.
Once data categorisation is complete, metadata is generated at this step, which will serve as a resource across all Makefiles within the pipeline.

\subsection{Dark signal removal} \label{sec:dark}

\begin{figure}
  \begin{center}
  \ifdefined\makepdf%
    \tikzsetnextfilename{figure-flowchart-pipeline}%
    \input{tex/src/figure-flowchart-pipeline.tex}%
  \else
    \includegraphics[width=0.99\linewidth]{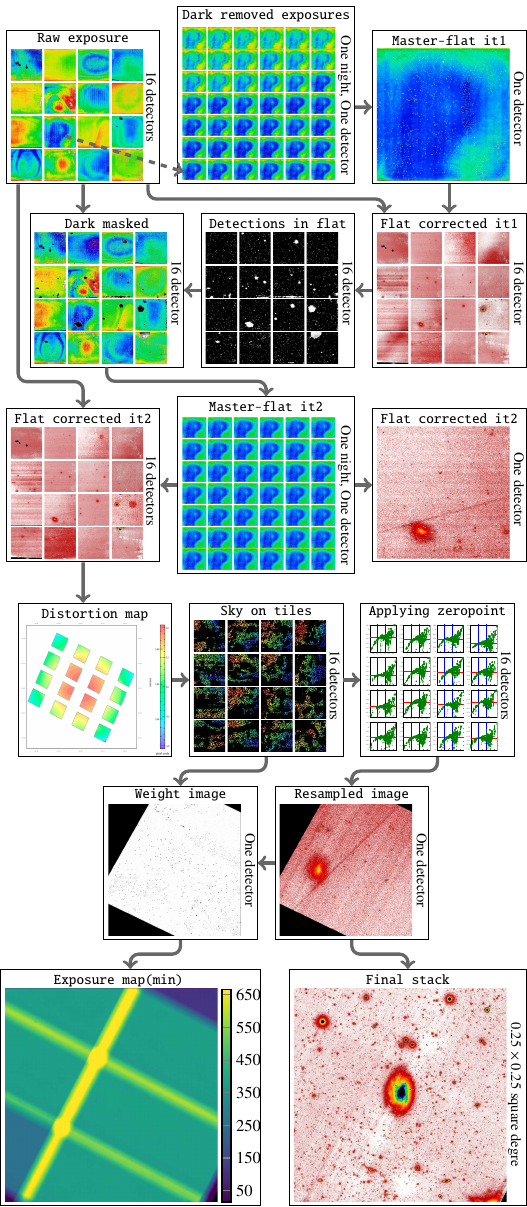}
  \fi

  \end{center}
  \vspace{-3mm}
  \caption{\label{fig:flowchart}
     Schematic overview of the NASIM pipeline, illustrating the principal stages of its VISTA/VIRCAM data reduction. Each panel represents a major processing step, including dark subtraction, flat-fielding, astrometric calibration, initial sky subtraction, weight estimation, photometric calibration, resampling, and final stacking.
  }
\end{figure}

In the KEDFS survey, each 1.5$\times$1 deg$^2$ tile is built up over approximately 20 nights to reach full depth, with a typical night contributing 42 exposures.
Given VIRCAM's detector configuration, each exposure involves 16 detectors.
Figure \ref{fig:flowchart} shows a flowchart outlining the major stages of data reduction in the NASIM pipeline.
It begins with a visual representation of raw VIRCAM images from all detectors associated with a specific exposure (\texttt{Raw exposure}), followed by dark-subtracted images for detector 7 from a single night (\texttt{Dark removed exposures}).
Removing dark frames results in a minimal reduction of the background, about $1\%$ in $K_s$-band images.
As a result, no discernible objects become visible even after subtracting the dark frames.
The dark frame removal process is carried out through the dark removal Makefile (\texttt{dark.mk}) in the NASIM pipeline, as shown in Figure \ref{fig:structure}.
Additionally, this Makefile also handles the process of using bad pixel masks to exclude defective pixels from the detectors.

\subsection{Sky flat fielding}

Accurate estimation of the flat-field correction is of paramount importance in studying LSB structures.
To gauge the relative sensitivity of the pixels, a uniform external light source is essential.
However, conventional methods such as dome flats or twilight flats are inadequate for our objectives due to imperfect flat illumination and varying sky levels over time.
A superior method involves using the science images themselves to create master flats \citep[e.g.][]{TrujilloFliri16, Trujillo21}.
This approach offers the distinct advantage of assessing the relative sensitivity among pixels at the identical intensity level found in the images.
The very bright NIR sky ($\mu_K \sim 14.85$ mag\,arcsec$^{-2}$) serves well as a uniform light source.
Nonetheless, challenges arise from diffuse objects and the extended point spread function, which introduce potential biases in the median of images.
These challenges can be mitigated by robustly masking astrophysical and optical signal, reducing bias in the background estimation.

In the NASIM pipeline, a two-iteration approach is used for flat-fielding.
In the initial stage of flat-fielding, the process begins by normalising all the dark-subtracted images.
Subsequently, using Gnuastro's \textsc{Arithmetic} program, we apply a $\sigma$-clipped median stacking method to combine the normalised science images for each night and detector.
An example of the master flat (\texttt{Master-flat it1}) is presented in Figure \ref{fig:flowchart}.
The science images are then divided by these first-step master flats (\texttt{Flat corrected it1}).
Once this flat is removed, objects become increasingly visible, allowing for improved masking in the next iteration of flat-fielding.

\subsubsection{Image masking}

To effectively mask the objects in the image, we employed the \textsc{NoiseChisel} program from Gnuastro \citep{gnuastro, Akhlaghi19a}.
\textsc{NoiseChisel} uses a non-parametric technique involving a threshold below the sky level and erosion for precise signal detection, which is particularly effective for objects with irregular shapes that are embedded within the noise.
After the initial round of flat correction, we used \textsc{NoiseChisel} in the NASIM pipeline to mask objects within the flat-corrected images.
Subsequently, we utilised Gnuastro's \textsc{Arithmetic} program to apply these masks to the dark-subtracted images, preparing them for inclusion in the second iteration of flat-fielding.
We illustrate a representative sample of detections for one exposure (\texttt{Detections in flat}), along with dark-removed masked images (\texttt{Dark masked}), in Figure \ref{fig:flowchart}.

\subsubsection{Sky variation with time}

\begin{figure*}[t]
  \begin{center}
  \ifdefined\makepdf%
    \tikzsetnextfilename{figure-flat}%
    \input{tex/src/figure-flat.tex}%
  \else
    \includegraphics[width=0.95\linewidth]{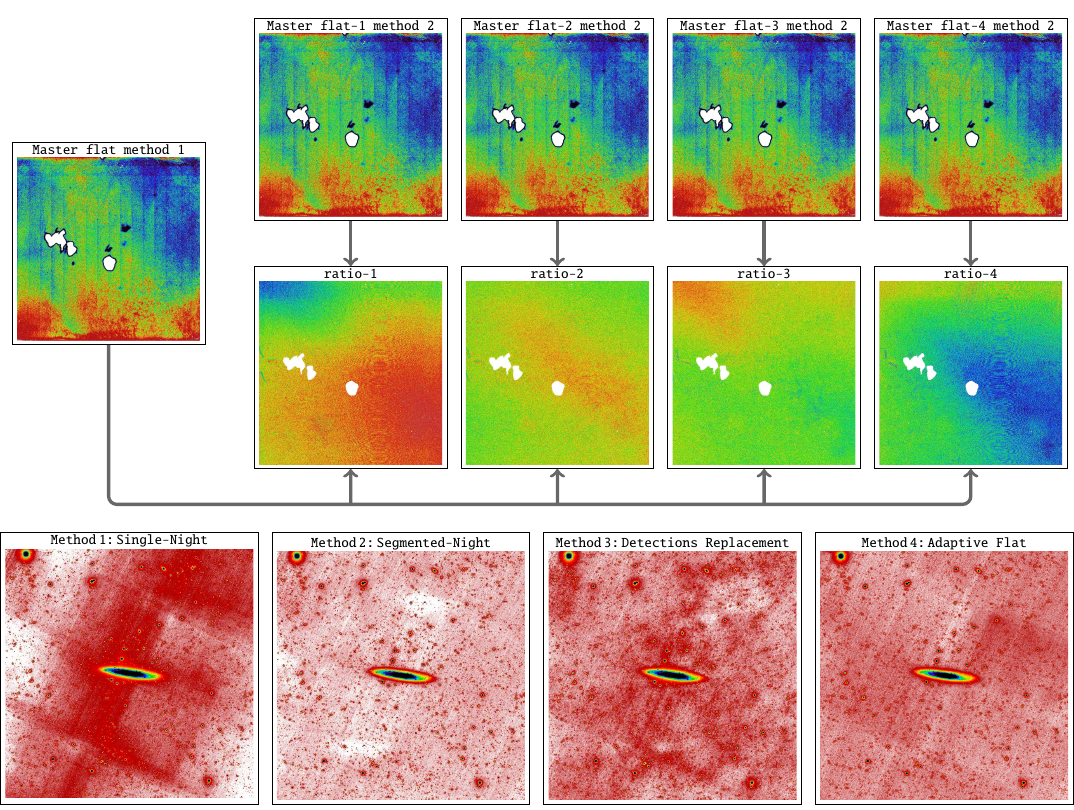}
  \fi

  \end{center}
  \vspace{-3mm}
  \caption{\label{fig:flat}
  Illustration of flat-fielding strategies and their outcomes in the NASIM pipeline.
Top panels: Ratios of segmented-night master flats (second method) to the single-night master flat (first method), revealing significant sky variations over approximately one hour within a single night. These fluctuations highlight the need for special flat-fielding techniques.
Bottom panels: Final coadded images using different flat-fielding methods, each covering 0.28$\times$ 0.28 deg$^2$ and centred on the galaxy IC\,2000 from KEDFS\_1\_5 tile: \texttt{Method\,1:\,Single-Night}, preserves galaxy outskirts but retains large-scale gradients; \texttt{Method\,2:\,Segmented-Night}, reduces gradients but suffers from persistence artefacts; \texttt{Method\,3:\,Detections Replacement}, attempts to mitigate persistence effects, but introduces background inconsistencies; \texttt{Method\,4:\,Adaptive Flat}, achieves the best balance, effectively reducing both gradients and persistence.
  }
\end{figure*}

During the second iteration of flat-fielding, we explored multiple strategies to achieve the optimal result given the variation in the sky level as a function of time.
This section outlines these techniques:

\begin{itemize}
\item
  Method 1: Single-night masterflats

  In this approach, we employed a methodology similar to that of the first flat-fielding.
Specifically, all the dark-masked images captured during a single night for each detector were utilised to generate a single masterflat.
Given the number of detectors, this resulted in 16 master flats per night.
A visual representation of this approach can be seen in Figure \ref{fig:flat} (\texttt{Method\,1:\,Single-Night}), which displays a coadded final image from the pipeline, featuring the galaxy IC\,2000 at the centre (from KEDFS\_1\_5 tile).
Covering an area of 0.28$\times$ 0.28 deg$^2$, this image was formed by combining data from 23 observation nights.
For a more comprehensive overview of the stacking process, refer to Section \ref{sec:stack}.
While this method effectively preserved the outskirts of the galaxy, we proceeded to explore alternative methods to mitigate background gradients without compromising the visibility of diffuse objects.

\vspace{2mm}
\item
  Method 2: Segmented-night masterflats

  To address variations in the sky within a single night, we adopted an alternative approach.
Instead of generating a master flat for each night and detector, we created a master flat for every sequential set of 10 or 11 images observed in succession.
Given that the typical number of images observed in a night is 42 in the KEDFS survey, this technique yields 4 master flats per night.
Figure \ref{fig:flat} highlights the rapid sky background variations observed within a single night, despite the exposures covering only about one hour.
The ratio of master flats in the second method, compared to a single master flat in the first method, clearly displays the significant variations in the sky.

Although this approach improves the gradients, it introduces a new complication.
Unfortunately, due to the absence of sufficient and proper dithering in the KEDFS observations, reducing the number of science images used to create the master flats led to the manifestation of the persistence effect for large objects.
Namely, the white ``holes'' caused by this effect can be seen in Figure \ref{fig:flat} (\texttt{Method\,2:\,Segmented-Night}).

\vspace{2mm}
\item
  Method 3: Detections replacement in masterflats

  In this method, we followed a procedure closely resembling that of Method 2 for creating master flats. However, before applying the master flats to the science images, we replaced the values in the second master flats with those of the first master flats in regions containing detections. Despite our efforts to mitigate the persistence effect, this adjustment resulted in an exacerbation of background gradients (see Figure \ref{fig:flat}, \texttt{Method\,3:\,Detections Replacement}).

  To improve the outcomes following the replacement of values in the second master flats with those from the first master flats in the regions containing detections, we developed a process involving the computation of ratios between the second and first master flats.
We then applied surface modelling to these ratios, allowing us to predict and replace the values of detection pixels (set to 1) with new values based on the model's projections.
These modified ratios were subsequently multiplied by the first master flats to generate adjusted second master flats, which we then applied to the science images.
However, the applied modifications did not lead to significant improvements in the outcomes.

\vspace{2mm}
\item
  Method 4: Adaptive flat

  In our final approach, we implemented a novel strategy.
This time, we chose to generate a unique master flat for each science image captured within a single night.
For example, if 42 images were taken during a single night, we created 42 individual master flats (as depicted in Figure \ref{fig:flowchart} by \texttt{Master-flat it2}).
Each master flat was tailored to its corresponding science image (image N) and constructed using data from 11 consecutive images.
This set included image N, along with 5 images taken before it and 5 images taken after it.
The intricate process was facilitated by the flat Makefile (see \texttt{flat.mk} in Figure \ref{fig:structure}).
The results of this method, as shown in the final coadd image in Figure \ref{fig:flat} (\texttt{Method\,4:\,Adaptive Flat}), were exceptionally promising.
We successfully reduced the persistence effect to an acceptable level, effectively mitigating background gradients.
This approach represents a significant advancement in our flat field correction methodology.
\end{itemize}

Before the second flat correction, we applied masking to the master flats to address specific problematic areas.
These problematic pixels were primarily located along the edges of the detectors, where pixel values deviated significantly from unity, potentially causing irregularities in the final stack.

\subsubsection{VIRCAM 1/f noise removal}

We identified subtle horizontal patterns in the final processed images, characterised by very small amplitude variations ($\sim0.05\%$).
Despite their minimal magnitude, these artefacts required correction prior to combining all exposures.
They appeared to be closely related to the camera’s readout process, prompting us to remove them by subtracting the excess counts from the images.

\begin{figure}
  \begin{center}
  \ifdefined\makepdf%
    \tikzsetnextfilename{figure-1f-noise}%
    \input{tex/src/figure-1f-noise.tex}%
  \else
    \includegraphics[width=0.99\linewidth]{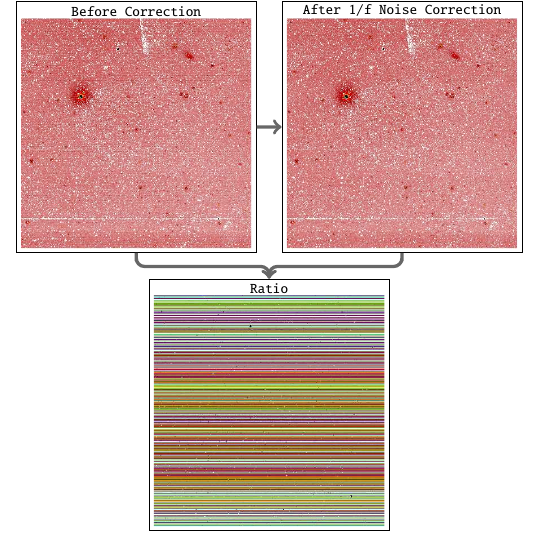}
  \fi

  \end{center}
  \vspace{-3mm}
  \caption{\label{fig:fnoise}
    Correction of 1/f noise artefacts in the final images. Top panels: Comparison between a sample image \texttt{Before Correction} (left) and the same image \texttt{After 1/f Noise Correction} (right) after correction using Gnuastro’s \textsc{Arithmetic} program. Bottom panel: The \texttt{Ratio} between the two images, highlighting the spatial structure of the removed pattern.
  }
\end{figure}

To implement this correction, we employed Gnuastro’s \textsc{Arithmetic} program.
First, we used the \texttt{collapse-mean} operation to reduce the dimensionality of the dataset, computing the mean value along the collapsed dimension.
This effectively removed one axis while retaining double-precision floating-point accuracy.
We then used the \texttt{add-dimension-fast} option to reconstruct a higher-dimensional dataset by stacking all input arrays consecutively along the fastest-varying dimension.
A visual demonstration of the effect of this correction is shown in Figure \ref{fig:fnoise}.
Although the 1/f noise removal significantly improves image quality, it does not completely eliminate all structured patterns.
These residual effects become noticeable in the final coadded image (see \texttt{Final stack} in Figure \ref{fig:flowchart}), and we are actively working on improving this step to achieve better results.

\subsection{Astrometric calibration}

We performed astrometric calibration in two stages within the astrometry Makefile (see \texttt{astrometry.mk} in Figure \ref{fig:structure}).
First, a rapid solution is derived using \textsc{Astrometry.net} \citep{Lang10} with reference to \textsc{Gaia} DR3 \citep{Gaia21}, providing an initial WCS alignment.
This solution is then refined using \textsc{SCAMP} \citep{Bertin06}, which corrects residual distortions and ensures precise cross-detector alignment using source catalogues from \textsc{SExtractor} \citep{Bertin96}.
For details of the procedure, configuration choices, and performance validation, see Appendix \ref{app:astrometry}.

\subsection{Sky subtraction} \label{sec:sky}

Accurate sky subtraction is critical for preserving the properties of LSB features in extended sources.
Aggressive background removal introduces significant over-subtraction resulting in negative fluxes around large galaxies, an effect observed in many surveys.
Although most evident near the largest sources, where they cover a substantial fraction of the field of view, this bias also affects the detection and photometry of unresolved sources across a wide range of redshifts.
In the NIR bands, the sky is substantially brighter than in the optical, with an average surface brightness of approximately 14.85 mag\,arcsec$^{-2}$ in the $K_s$ band.
This value is reported as $\mu_K \sim 13$ mag arcsec$^{-2}$ in \citet{Cuby00}, though the magnitude system is not explicitly specified. We assume it is in the Vega system and convert it to the AB system using an offset of 1.85 mag for the $K_s$ band \citep{Blanton07}.
The extensive and diffuse outskirts of large galaxies can become indistinguishable from the sky, and local interpolation techniques risk subtracting these features as part of the sky, leading to their loss in the final stacked images.

To mitigate this, we adopted a less aggressive sky subtraction approach, in which we estimated and subtracted a single background value from each frame, avoiding explicit sky modelling.
We estimated sky values using the \textsc{NoiseChisel} program.
To enhance the signal-to-noise ratio during this estimation, we used tiles rather than individual pixels.
We strengthened outlier rejection by increasing the \texttt{interpnumngb} parameter, and enforced tile purity by setting \texttt{minskyfrac} to 0.9—selecting only tiles in which at least 90\% of the area was free from detected astronomical sources.
A representative example of the sky image for a single exposure (Sky on tiles) is shown in Figure \ref{fig:flowchart}.
For each image, we computed and subtracted the $\sigma$-clipped median of the resulting sky image, using $3\sigma$ clipping with a tolerance of 0.1.
The entire sky subtraction process is managed through the sky Makefile (\texttt{sky.mk}) within the NASIM pipeline, as depicted in Figure \ref{fig:structure}.

\subsection{Image weighting} \label{sec:weight}

To account for variations in image quality due to airmass changes and detector-specific effects, NASIM assigns relative weights to individual exposures using the standard deviation of the sky (SKY-STD) as a proxy for noise.
Weights are normalised such that the best-quality images receive a value of 1.0, and poorer-quality frames are scaled accordingly.
This process is managed via the \texttt{weight.mk} Makefile (see Figure \ref{fig:structure}).
Full details of the weighting procedure are provided in Appendix \ref{app:weighting}, including an example of SKY-STD variation across detectors and its correlation with airmass.

\subsection{Photometric calibration}

Photometric calibration in NASIM is based on the 2MASS catalogue, with magnitudes converted to the AB system \citep{Blanton07}.
All calibration steps are orchestrated within a dedicated Makefile (\texttt{zeropoint.mk}), as shown in Figure \ref{fig:structure}.
We performed stellar photometry using Gnuastro tools and estimated the photometric zero points following the method implemented in the \textsc{astscript-zeropoint} program \citep{Eskandarlou23}.
Zero points are derived using an optimal aperture radius of 2 arcsec and normalised to a constant value of 27 mag.
A representative example of the magnitude difference between 2MASS and KEDFS sources—measured across 42 images from a single night and across different detectors—is illustrated in the \texttt{Applying zeropoint} panel of Figure \ref{fig:flowchart}.
Full details of the calibration procedure are provided in Appendix \ref{app:photometry}.

\subsection{Resampling} \label{sec:resamp}

We used Gnuastro’s \textsc{Warp} program to resample images onto a common grid aligned with WCS coordinates.
We adopted a pixel scale of 0.2 arcsec to improve PSF sampling and enable robust measurement of LSB features.
To optimise memory and computational resources, we restricted resampling to subregions of $\sim$0.3 deg$^2$ centred on the target.
We coordinated this step via the \texttt{resample.mk} Makefile (see Figure \ref{fig:structure}).
After resampling, we applied scalar image weights (Section \ref{sec:weight}) directly.
Representative outputs, including resampled science and weight images, are shown in Figure \ref{fig:flowchart}.
Full technical details are provided in Appendix \ref{app:resampling}.

\begin{figure*}
  \begin{center}
  \ifdefined\makepdf%
    \tikzsetnextfilename{figure-loopflow}%
    \input{tex/src/figure-loopflow.tex}%
  \else
    \includegraphics[width=0.95\linewidth]{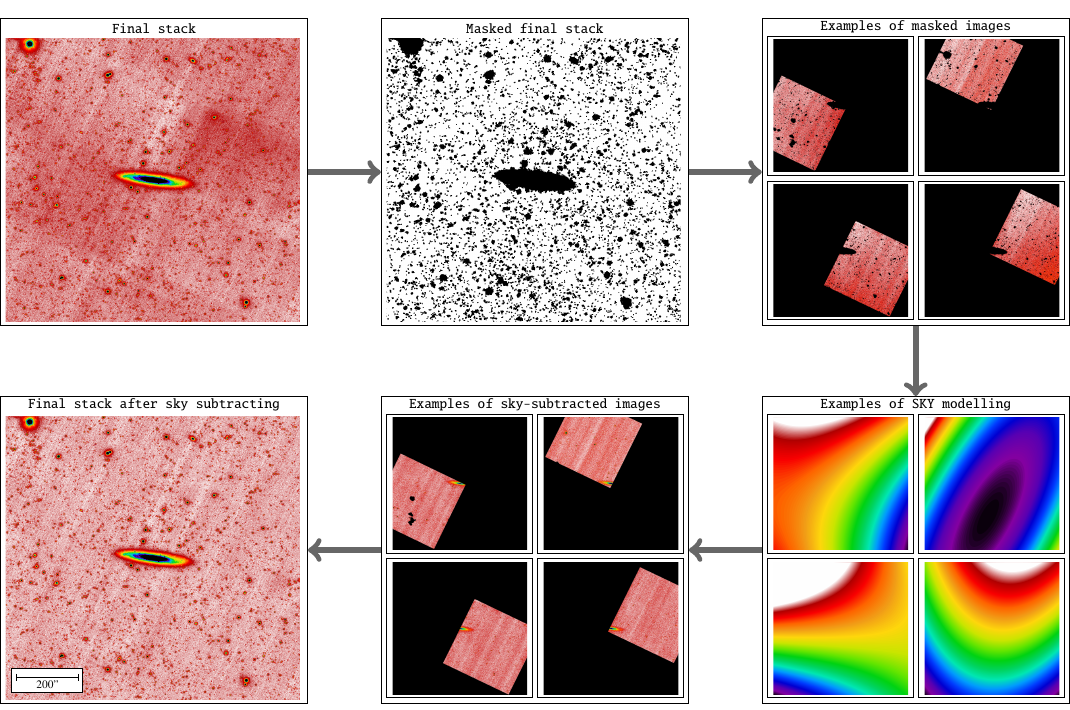}
  \fi

  \end{center}
  \vspace{-3mm}
  \caption{\label{fig:loopflow}
    Sky modelling and subtraction process in the NASIM pipeline, illustrated for the IC\,2000 galaxy. This figure shows the main steps within the sky subtraction loop implemented via the \texttt{loopflow.mk} Makefile. Top-left: \texttt{Final stack} image produced by the NASIM pipeline prior to additional sky modelling. Top-middle: \texttt{Masked final stack}, created by applying a strong mask with \textsc{NoiseChisel} to isolate background regions. Top-right: \texttt{Examples of masked images}, showing individual masked, resampled frames. Bottom-right: \texttt{Examples of sky modelling}, based on fits with a second-order polynomial function. Bottom-middle: \texttt{Examples of sky-subtracted images}, following subtraction of the sky model. Bottom-left: \texttt{Final stack after sky subtraction}, demonstrating improved background uniformity and enhanced recovery of LSB features.
  }
\end{figure*}

\subsection{Final stack} \label{sec:stack}

Prior to co-adding the science images, we had to identify and mask artefacts such as saturated pixels, streaks, and cosmic rays at the pixel level.
And then we employed a $\sigma$-clipping procedure for both the median and standard deviation (STD) using Gnuastro's \textsc{Arithmetic} program.
Pixels with values exceeding or falling below \( 3 \times \text{STD} + \text{median} \) were masked to ensure that such artefacts were excluded from the final combination.
The same masking procedure was subsequently applied to the weight images generated during the resampling stage (see Section \ref{sec:resamp}).

The final stack was produced by dividing the weighted sum of the images by the sum of their corresponding weights, according to the following relation:

\begin{equation}
\text{Stacked image} = \frac{\sum_{i=1}^{N} w_{i} I_{i}}{\sum_{i=1}^{N} w_{i}},
\label{eq:stacking}
\end{equation}
where \( I_{i} \) and \( w_{i} \) denote the intensity and weight of the \(i\)-th image, respectively.

All these steps were executed within a single Makefile, \texttt{stack.mk}, as illustrated in Figure \ref{fig:structure}.
In addition to producing the final stacked image and associated weight map, this Makefile also generates an exposure map based on the number of contributing images.
Since each exposure has an integration time of one minute, the exposure map is expressed in minutes.

Examples of the final stacked image (\texttt{Final stack}) and the corresponding exposure map (\texttt{Exposure map}) are shown in Figure \ref{fig:flowchart}.
For this illustration, we selected a 0.25 square degree region centred on the galaxy NGC\,1494, located within the KEDFS\_2\_4 tile of the KEDFS survey.
The exposure map indicates that the number of stacked images varies from approximately 40 to 660 across the field, with the majority of the area reaching a depth of around 5 hours.
A more detailed discussion of this image, focusing on NGC\,1494, is presented in Section \ref{sec:galaxy}.

\begin{figure*}
  \begin{center}
  \ifdefined\makepdf%
    \tikzsetnextfilename{figure-ngc1494}%
    \input{tex/src/figure-ngc1494.tex}%
  \else
    \includegraphics[width=0.99\linewidth]{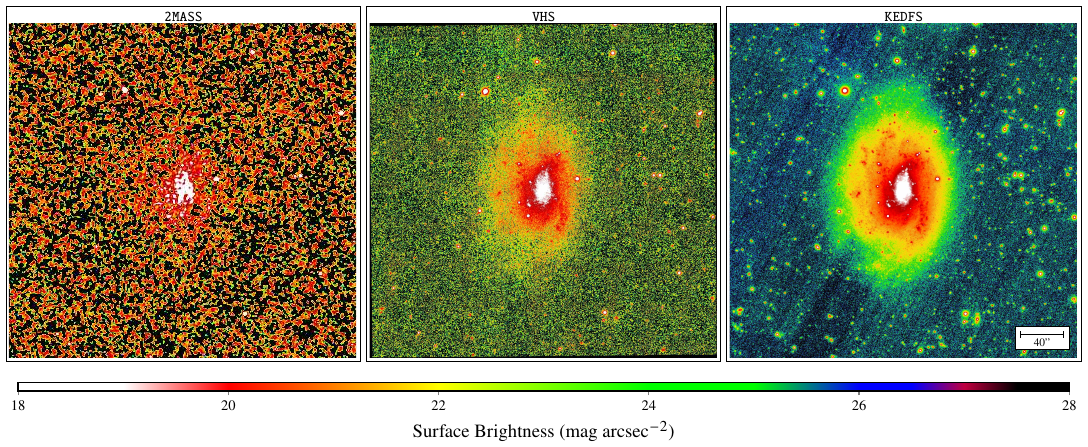}
  \fi

  \end{center}
  \vspace{-3mm}
  \caption{\label{fig:ngc1494}
  Comparison of $K_s$-band images of NGC\,1494 from 2MASS (left), VHS (middle), and KEDFS (right). All panels are shown with the same physical scale and intensity stretch to emphasise faint structures. The KEDFS image, processed with the NASIM pipeline, reveals a markedly more extended outskirts and LSB features that are not detected in the shallower 2MASS and VHS data.
  }
\end{figure*}

\subsection{Sky modelling} \label{sec:loopflow}

Following the creation of the final stacked image, we applied an additional stage to model and subtract the residual sky using a strong mask.
This procedure was implemented through a new Makefile in the NASIM pipeline, referred to as \texttt{loopflow.mk}.
An overview of this process, together with illustrative examples, is presented in Figure \ref{fig:loopflow}.

Initially, we employed Gnuastro's \textsc{NoiseChisel} program to generate a strong mask based on the final stack.
Subsequently, using Gnuastro's \textsc{Warp} and \textsc{Arithmetic} programs, this mask was propagated to each individual resampled image.
For the modelling of the residual sky, we tested three approaches: fitting a two-dimensional Chebyshev polynomial, and first- and second-order polynomial functions.
After evaluating the performance of each method, we adopted the second-order polynomial function for use in the NASIM pipeline.

Following this additional sky subtraction, several steps of the pipeline were repeated, including the estimation of image weights, the construction of weight images, the identification of outliers, and the final stacking.
This iterative nature of the process motivated the designation of the Makefile as \texttt{loopflow.mk}.
As illustrated in Figure \ref{fig:loopflow}, the final stacked image produced by this procedure demonstrates a significant improvement in sky subtraction, while effectively preserving the LSB structures.

\subsection{Efficient processing}

The NASIM pipeline provides a fully automated, scalable, and reproducible framework for the reduction of VISTA/VIRCAM data.
By employing 50 CPU cores in parallel, it completes the full workflow, from initial data retrieval from the ESO archive to final image stacking, in under 5 seconds per frame.
Designed with large-scale surveys in mind, NASIM significantly reduces computational demands, lowering memory usage by 2.86\,GB per image.
Across the entire KEDFS dataset, this optimisation results in a cumulative storage saving of approximately 755\,TB.
The pipeline is inherently flexible, allowing seamless application to any VISTA/VIRCAM survey, independent of filter selection or survey depth.
A demonstration of its performance, based on a comparison between NASIM-processed and standard publicly released data from the VIDEO survey, is presented in Section \ref{sec:discussion}.

\section{Science showcases} \label{sec:showcase}

In this section, we present several scientific cases enabled by the NASIM pipeline applied to the KEDFS survey data.
These examples not only demonstrate the pipeline’s ability to process complex, large-scale datasets, but also highlight its effectiveness in recovering LSB structures and detecting faint sources in the NIR.
In addition, we introduce the depth and key parameters of this new VISTA survey, previously unreleased, providing unique opportunities for future investigations.

\begin{figure}
  \begin{center}
  \ifdefined\makepdf%
    \tikzsetnextfilename{figure-sb-ngc1494}%
%
%

\begin{tikzpicture}

  \usetikzlibrary {decorations.text}

  \scriptsize

  \draw [white](-1cm,-0.8cm) rectangle (8cm,7.4cm);

    \begin{axis}[
        xlabel={R (arcsec)},
        ylabel={$\mu_{\mathrm{K_s}}$ (mag arcsec$^{-2}$)},
        legend pos=outer north east,
        width=9.3cm,
        height=8.8cm,
        ymin=19.0,
        ymax=29.9,
        y dir=reverse,
        xmin=-5, xmax=150,
        xtick={0,20,40,60,80,100,120,140},
        xticklabels={0, 20, 40, 60, 80, 100, 120, 140},
        grid=both,
        grid style={line width=.1pt, draw=gray!20},
        major grid style={line width=.2pt, draw=gray!20},
        legend style={
            at={(0.85,0.96)}, 
            anchor=north, 
            draw=black, 
            fill=white, 
            text height=1.5ex, 
            text depth=.25ex, 
            column sep=1ex, 
        }
    ]

    \addplot[
        color=purple,
        mark=square*,
        mark size=1.4pt,
        only marks,
        restrict y to domain=0:30,
        error bars/.cd,
        y dir=both,
        y explicit,
    ] table[x index=0, y index=2, y error index=3, col sep=space]
            {tex/build/texts/sb-k-nasim-ngc1494.txt};
    \addlegendentry{KEDFS}

    \addplot[
        color=teal,
        mark=*,
        mark size=1.5pt,
        only marks,
        restrict y to domain=0:30,
        error bars/.cd,
        y dir=both,
        y explicit,
    ] table[x index=0, y index=2, y error index=3, col sep=space]
            {tex/build/texts/sb-k-vhs-ngc1494.txt};
    \addlegendentry{VHS}

    \addplot[
        color=violet,
        mark=triangle*,
        mark size=1.5pt,
        only marks,
        restrict y to domain=0:29,
        error bars/.cd,
        y dir=both,
        y explicit,
    ] table[x index=0, y index=2, y error index=3, col sep=space]
            {tex/build/texts/sb-k-2mass-ngc1494.txt};
    \addlegendentry{2MASS}

    \end{axis}


\end{tikzpicture}%
  \else
    \includegraphics[width=0.99\linewidth]{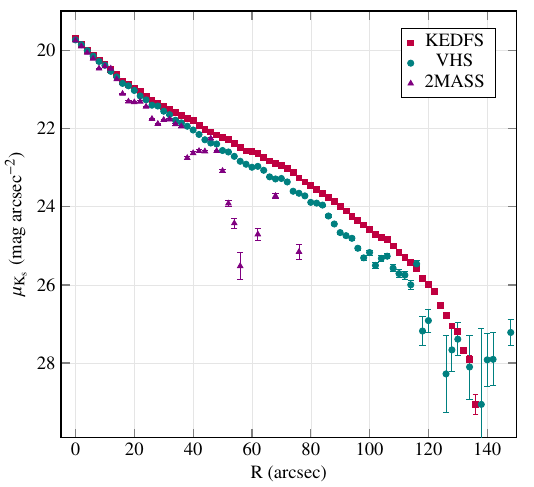}
  \fi

  \end{center}
  \vspace{-3mm}
  \caption{\label{fig:sbngc1494}
    Radial surface brightness profiles of NGC\,1494 in the $K_s$ band, derived from 2MASS, VHS, and KEDFS images using the Gnuastro \textsc{astscript-radial-profile} script. The KEDFS profile extends smoothly to fainter levels (reaching $\sim$27.7 mag\,arcsec$^{-2}$), demonstrating both the superior depth of the survey and the effectiveness of the NASIM pipeline in preserving LSB features in the galaxy's outskirts.
  }
\end{figure}
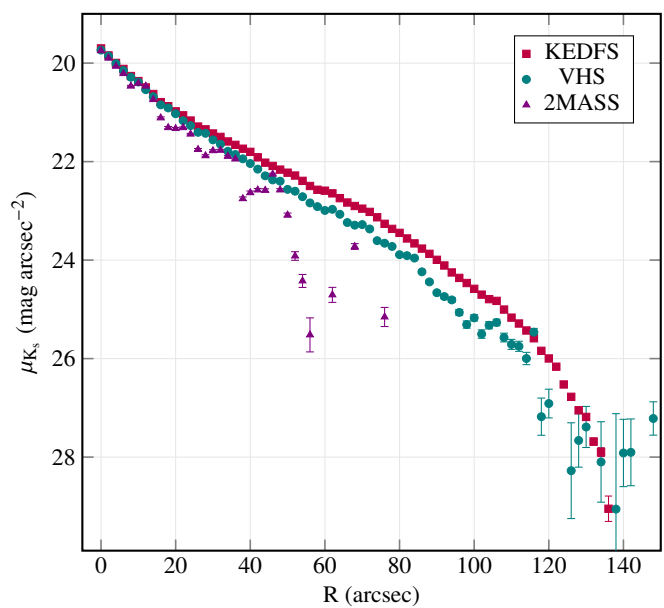

\subsection{Outskirts and edge of galaxies} \label{sec:galaxy}

One of the key strengths of the NASIM pipeline is its ability to reveal LSB features in the outskirts of galaxies, regions critical for studying galaxy growth and tracing the full extent of galactic structures.
Here, we examine the galaxy NGC\,1494 (also shown in the pipeline flowchart; Figure \ref{fig:flowchart}) as a case study to demonstrate both the depth achieved by the KEDFS survey and the effectiveness of the NASIM pipeline in preserving faint, extended structures.

NGC\,1494 is a late-type spiral galaxy \citep{Grosbol04}, classified as SA(s)d \citep{Buta15}, located at R.A.(J2000) = 03$^{\mathrm{h}}$57$^{\mathrm{m}}$43$^{\mathrm{s}}$.7, Dec(J2000) = $-48^\circ$54$'$22$''$, at a distance of approximately 13.8\,Mpc \citep{Pisano11}.
Its moderately inclined disc and relatively undisturbed morphology make it an excellent target for investigating the faint outskirts of galaxies.

Figure \ref{fig:ngc1494} presents images of NGC\,1494 from three datasets of varying depth: our NASIM-processed reduction of the KEDFS field, the VHS survey processed by the CASU pipeline, and 2MASS.
This visual comparison highlights both the superior depth of the KEDFS survey and the capability of the NASIM pipeline to preserve LSB structures, particularly in the galaxy’s outer regions where signal levels approach the background noise.

To quantitatively support the visual differences seen in Figure \ref{fig:ngc1494}, we derive $K_s$-band surface brightness profiles for NGC\,1494 using the 2MASS, VHS, and KEDFS datasets.
We compute these profiles, shown in Figure \ref{fig:sbngc1494}, using the \textsc{astscript-radial-profile} tool from Gnuastro \citep{infantesainz24b}.
This script constructs elliptical annuli and measures the flux distribution using routines from the Gnuastro suite.
Before extracting the profiles, we mask all surrounding sources using \textsc{NoiseChisel} to minimise contamination.
For NGC\,1494, we adopt elliptical apertures with an axis ratio of 0.65 and a position angle of 85.75° (measured east of north) to accurately trace the galaxy’s structure.

The KEDFS reduction reaches a $K_s$-band surface brightness depth of approximately 27.7\,mag\,arcsec$^{-2}$ (3$\sigma$ over 100\,arcsec$^2$), corresponding to roughly 67 times higher sensitivity than 2MASS and 11 times higher than VHS in terms of surface brightness detection.

\subsection{Low surface brightness galaxies}

A class of galaxies known as LSB galaxies has been studied for decades \citep[e.g.][]{Sandage84, Caldwell87, Impey88}, particularly in dense environments such as galaxy clusters \citep{Conselice18}.
A subset of these, later referred to as ultra-diffuse galaxies (UDGs), was highlighted by \citet{vanDokkum15}, who identified many more faint examples and helped renew interest in this population.
With central surface brightnesses of $\mu_g \sim 24$--$26$ mag\,arcsec$^{-2}$ and effective radii of $r_\mathrm{e} \sim 1.5$--$4.5$~kpc, UDGs are considered promising candidates for probing the properties of nearly dark matter–dominated systems.
Progress in understanding UDGs has been hindered by several factors, including the lack of homogeneous observational datasets across environments, inconsistent and often arbitrary selection criteria, and limited constraints on the nature of their dark matter haloes \citep{Zaritsky23}.
Although many UDGs have been discovered in rich environments such as galaxy clusters, and field and group UDGs, may have distinct origins and evolutionary paths, offering unique constraints on the low-mass end of galaxy formation \citep{Mihos24}.

While most UDG discoveries have relied on deep optical imaging \citep[e.g.][]{Mihos15, Venhola17, Zaritsky19, Trujillo21}, NIR observations are essential for tracing their underlying stellar mass with reduced sensitivity to dust and young stellar populations.
The $K_s$ band, in particular, provides a powerful view of the stellar mass content of galaxies.
However, its potential for UDG studies remains largely untapped due to the high and variable NIR sky and the absence of specialised pipelines capable of recovering extremely faint, extended emission.
These challenges significantly limit the study of LSB galaxies in the NIR from ground-based observatories.
With NASIM, we address this gap by providing the first pipeline specifically optimised to recover LSB features in ground-based NIR data across all VISTA bands, with particular emphasis on the challenging $K_s$ band.

\begin{figure}
  \begin{center}
  \ifdefined\makepdf%
    \tikzsetnextfilename{figure-udg}%
    \input{tex/src/figure-udg.tex}%
  \else
    \includegraphics[width=0.95\linewidth]{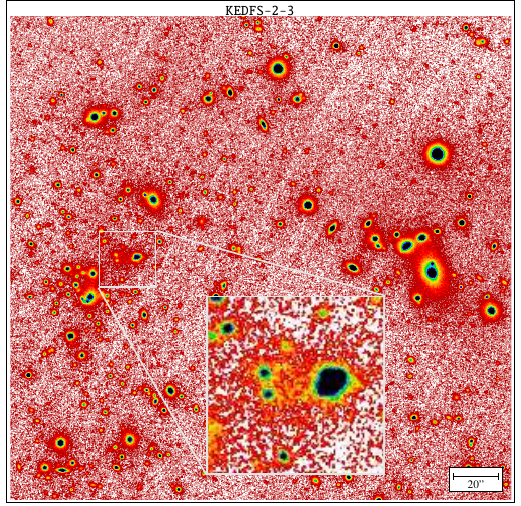}
  \fi

  \end{center}
  \vspace{-3mm}
  \caption{\label{fig:udg}
  Example of a UDG, \mbox{SMDG,J0406539--480442}, identified in the KEDFS field using the NASIM pipeline. The $K_s$-band image highlights the galaxy’s diffuse structure, with a zoom-in panel providing a closer view. Several galaxy clusters are also visible in the field, making this a particularly rich and interesting region. This figure illustrates NASIM’s ability to recover faint, extended features in crowded NIR fields.
  }
\end{figure}

In Figure \ref{fig:udg}, we present an example of a UDG identified in the KEDFS\_2\_3 tile: \mbox{SMDG\,0406539--480442}, a galaxy from the Systematically Measuring Ultra-Diffuse Galaxies (SMUDGes) catalogue \citep{Zaritsky19}.
It is located at R.A.(J2000) = 04$^\mathrm{h}$06$^\mathrm{m}$53$^\mathrm{s}$.91, Dec(J2000) = $-48^\circ$04$'$41$''$.7, and has a non-circularised effective radius of $8.28$ arcseconds based on Legacy Survey measurements \citep{Zaritsky22}.
Its detection in our $K_s$-band stack, despite the challenges posed by faint and extended surface brightness, clearly demonstrates NASIM’s ability to recover diffuse structures directly in the NIR.
This highlights the pipeline’s suitability for identifying and analysing LSB galaxies such as UDGs.
Building on this, future work will use NASIM-processed VISTA data alongside multi-wavelength observations to derive the physical and structural properties of UDGs across a range of environments.

\subsection{Intracluster light}

The diffuse stellar component known as ICL permeates galaxy clusters, composed of stars stripped from their host galaxies during tidal interactions, mergers, and other dynamical processes \citep[e.g.][]{MontesTrujillo14,Contini21,Montes22}.
As an unbound and extended tracer of past gravitational encounters, the ICL encodes the hierarchical growth history of clusters, offering a unique window into their assembly and evolution.
It serves as a sensitive probe of cluster dynamics, stellar mass distribution, and galaxy–environment interactions \citep[e.g.][]{DeMaio18,Montes19}.
Over the past two decades, deep optical surveys have firmly established the ubiquity of the ICL in both nearby and intermediate-redshift clusters, revealing spatial and colour gradients that trace the metallicity and age distribution of its stellar populations \citep[e.g.][]{Mihos17,Kluge20}.
However, the reliance on optical wavelengths introduces biases due to sensitivity to recent star formation, dust extinction, and redshift-dependent passband shifts—limitations that can be mitigated with deep NIR observations.

Deep infrared imaging from JWST has demonstrated the power of NIR data in revealing the structure and stellar populations of the ICL with robust clarity \citep{MontesTrujillo22}.
Meanwhile, recent studies using Euclid-like bands have begun to extend ICL analyses into the NIR from space, benefiting from stable backgrounds and improved spatial resolution \citep{Ellien25}.
However, Euclid lacks coverage in the $K_s$ band, which is particularly well suited for tracing the evolved stellar populations that dominate the ICL.
In this context, the KEDFS survey offers a crucial complement with the deepest ground-based $K_s$-band imaging in this region, enabling a more complete and less biased reconstruction of the ICL’s stellar mass and formation history.

\begin{figure}
  \begin{center}
  \ifdefined\makepdf%
    \tikzsetnextfilename{figure-gc}%
    \input{tex/src/figure-gc.tex}%
  \else
    \includegraphics[width=0.95\linewidth]{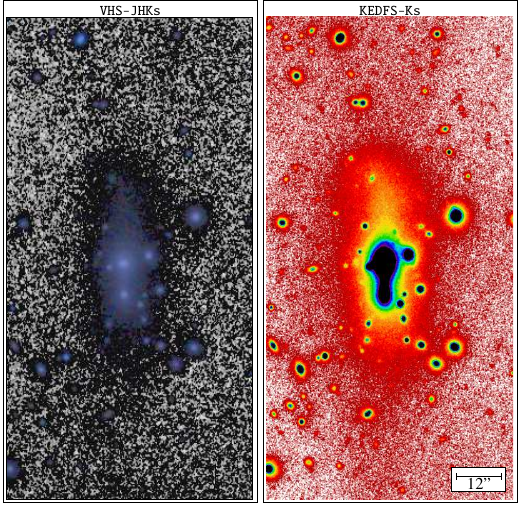}
  \fi

  \end{center}
  \vspace{-3mm}
  \caption{\label{fig:gc}
    Colour composite ($JHK_s$) image from VHS (left) compared to the deeper KEDFS $K_s$-band image processed with NASIM (right) for the galaxy cluster ACT-CL J0414.2$-$4612. The NASIM reduction reveals extended ICL around the central galaxy that is not visible in the shallower VHS data.
  }
\end{figure}

To demonstrate the capabilities of NASIM in recovering diffuse intracluster emission, we present an example in the galaxy cluster ACT-CLJ0414.2$-$4612, located at R.A.(J2000) = 04$^{\mathrm{h}}$14$^{\mathrm{m}}$13$^{\mathrm{s}}$.0, Dec(J2000) = $-46^\circ$12$'$00$''$.1 with a photometric redshift of $z = 0.1894$ \citep{Hilton21}.
This cluster lies within the KEDFS\_1\_1 tile, where the depth of the $K_s$-band imaging enables the ICL to be distinctly traced over extended regions.
In Figure \ref{fig:gc}, we compare the NASIM-reduced $K_s$-band image of this cluster with a colour composite (JHK) generated from the shallower VHS data.
We produce the colour image using the \textsc{astscript-color-faint-gray} script from Gnuastro, which applies a non-linear transformation to compress the dynamic range of bright sources, enhancing the visibility of faint diffuse structures in the final composite \citep{infantesainz24a}.
This comparison highlights the dramatic improvement achieved through NASIM processing: we clearly reveal the ICL as a diffuse, spatially coherent structure surrounding the brightest cluster galaxy—features that remain barely visible or entirely absent in the VHS composite.
The depth and uniformity of the KEDFS $K_s$-band data, enhanced by NASIM, provide a unique opportunity to explore the ICL in unprecedented detail.
We are thus able to make robust measurements of its stellar content, morphology, and spatial extent, as well as its connection to the total mass distribution of the cluster.

\section {Discussion} \label{sec:discussion}

\subsection{An improved LSB reduction scheme}

\begin{figure*}
  \begin{center}
  \ifdefined\makepdf%
    \tikzsetnextfilename{figure-nasim-casu}%
    \input{tex/src/figure-nasim-casu.tex}%
  \else
    \includegraphics[width=0.95\linewidth]{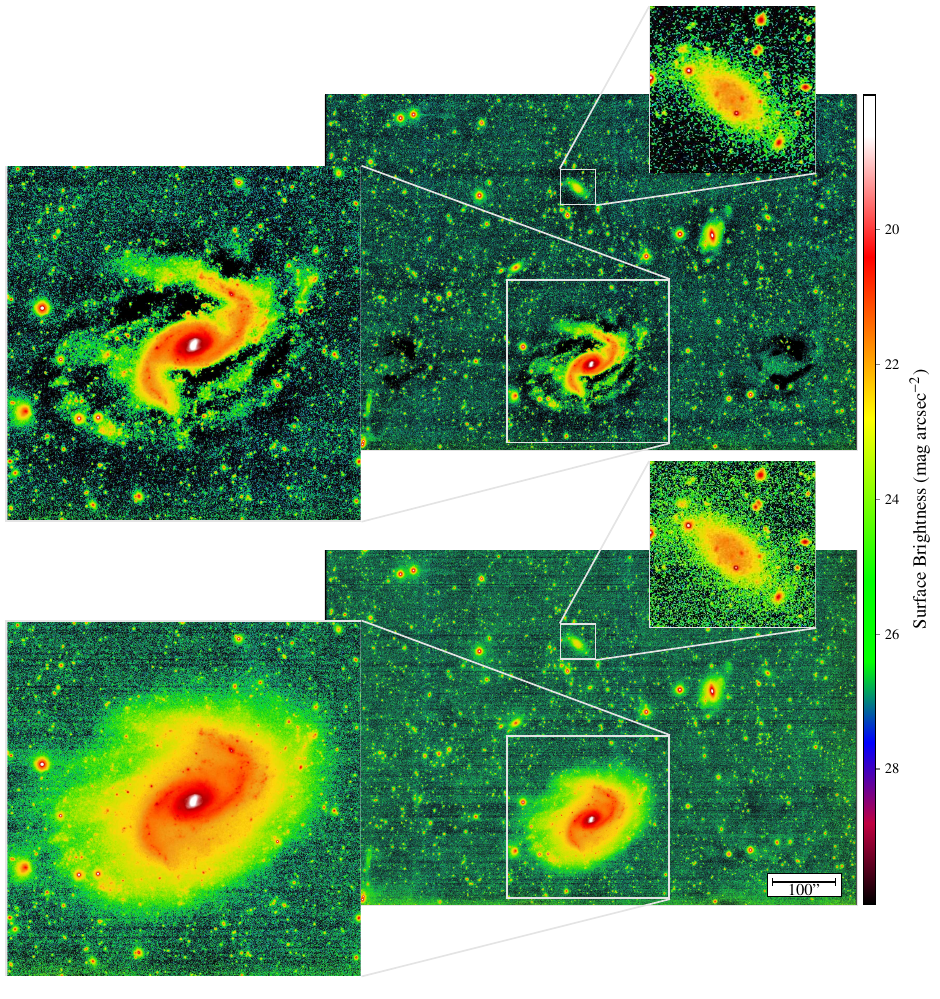}
  \fi

  \end{center}
  \vspace{-3mm}
  \caption{\label{fig:nasim-casu}
  Comparison between the standard VDFS-reduced VIDEO image (top) and the NASIM-processed version (bottom) for the field containing NGC\,895. The panels reveal a clear improvement in background uniformity and the recovery of extended LSB features in the NASIM reduction. The outer regions of NGC,895, which appear truncated in the VDFS image due to over-subtraction, are fully recovered. Ghost-like residuals seen in the standard reduction—likely caused by persistence effects from the bright galaxy, are absent in the NASIM result, demonstrating the effectiveness of the adaptive flat-fielding approach. Although the background is not perfectly uniform due to the original observing strategy, which was not optimised for NASIM’s flat-fielding method, significant improvements are still evident. Zoom-in panels highlight differences in the treatment of both the main galaxy and a nearby dwarf galaxy (LEDA 3098171).
  }
\end{figure*}

\begin{figure}
  \begin{center}
  \ifdefined\makepdf%
    \tikzsetnextfilename{figure-sb-ngc895}%
%
%

\begin{tikzpicture}

  \usetikzlibrary {decorations.text}

  \scriptsize

  \draw [white](-1cm,-0.8cm) rectangle (8cm,7.4cm);

    \begin{axis}[
        xlabel={R (arcsec)},
        ylabel={$\mu_{\mathrm{K_s}}$ (mag arcsec$^{-2}$)},
        legend pos=outer north east,
        width=9.3cm,
        height=8.8cm,
        ymin=17,
        ymax=29.9,
        y dir=reverse,
        xmin=-5, xmax=193.5,
        xtick={0,25,50,75,100,125,150,175},
        xticklabels={0,25,50,75,100,125,150,175},
        grid=both,
        grid style={line width=.1pt, draw=gray!20},
        major grid style={line width=.2pt, draw=gray!20},
        legend style={
            at={(0.77,0.96)}, 
            anchor=north, 
            draw=black, 
            fill=white, 
            text height=1.5ex, 
            text depth=.25ex, 
            column sep=1ex, 
        }
    ]

    \addplot[
        color=purple,
        mark=square*,
        mark size=1.4pt,
        only marks,
        restrict y to domain=0:30,
        error bars/.cd,
        y dir=both,
        y explicit,
    ] table[x index=0, y index=2, y error index=3, col sep=space]
            {tex/build/texts/sb-k-nasim-ngc895.txt};
    \addlegendentry{NASIM Pipeline}

    \addplot[
        color=teal,
        mark=*,
        mark size=1.5pt,
        only marks,
        restrict y to domain=0:29.8,
        error bars/.cd,
        y dir=both,
        y explicit,
    ] table[x index=0, y index=2, y error index=3, col sep=space]
            {tex/build/texts/sb-k-casu-ngc895.txt};
    \addlegendentry{Standard Pipeline}

    \end{axis}


\end{tikzpicture}%
  \else
    \includegraphics[width=0.99\linewidth]{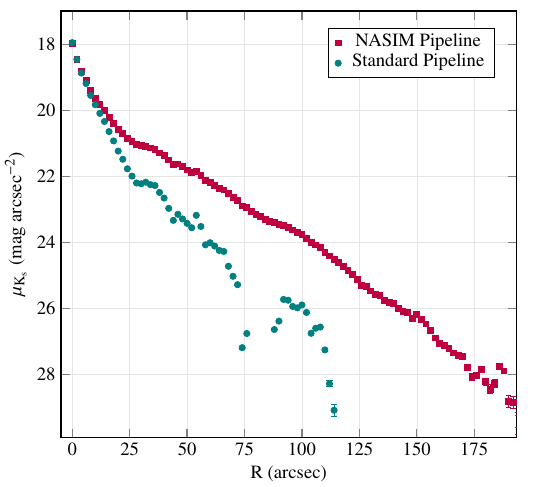}
  \fi

  \end{center}
  \vspace{-3mm}
  \caption{\label{fig:sbngc895}
  Surface brightness profiles of NGC\,895 extracted from the standard VIDEO reduction and the NASIM-processed image. The NASIM profile extends smoothly to larger radii and fainter surface brightness levels, demonstrating improved recovery of the galaxy’s outskirts. In contrast, the VIDEO reduction shows an artificial truncation beyond $\sim$10 arcsec due to over-subtraction of the sky.
  }
\end{figure}

The accurate recovery of extended LSB features in NIR imaging remains a significant challenge, particularly in wide-field surveys where conventional data reduction pipelines focus primarily on compact sources and point-source sensitivity in background-limited regimes.
These pipelines often apply aggressive sky subtraction and default flat-fielding procedures that, while optimised for point sources, inadvertently suppress or distort extended emission—especially in the outskirts of nearby galaxies or around diffuse intracluster structures.
In contrast, with NASIM, we specifically aim to preserve faint, extended LSB features.
We incorporate a suite of carefully tailored processing steps designed to overcome the principal limitations of conventional reductions:

\begin{itemize}

\item
 We apply a novel flat-fielding approach, known as adaptive flat, which constructs calibration frames directly from science exposures, using varying input sets tailored to produce a unique master flat for each image.
We implement this method using the powerful Gnuastro programs, which effectively remove large-scale instrumental patterns and correct for the complex detector-level structures of the VIRCAM array.

\vspace{2mm}
\item
  We apply a conservative sky subtraction strategy that avoids overfitting large-scale background variations and preserves genuine diffuse emission.
We complement this with optimised object masking using the robust \textsc{NoiseChisel} program and careful sky modelling on appropriate spatial scales to ensure an unbiased reconstruction of faint extended structures.

\vspace{2mm}
\item
  We perform robust outlier rejection during stacking and apply a weighted stacking scheme that accounts for image quality and background noise.
This strategy suppresses residual artefacts and transient features while preserving extended LSB signals across multiple exposures.

\end{itemize}

\subsection {Detailed comparison with point source-optimised pipeline}

To illustrate the practical advantages of the LSB-optimised algorithms of NASIM, we make a detailed comparison with the results from a point source-optimised pipeline applied to imaging of a field selected from the VISTA Deep Extragalactic Observations (VIDEO) survey.
As summarised in Table \ref{tab:vista}, VIDEO is a deep, multi-band programme covering approximately 12 square degrees across the $Z$, $Y$, $J$, $H$, and $K_s$ bands, designed to trace galaxy and large-scale structure evolution out to $z \sim 4$ \citep{Jarvis13}.
Its combination of depth and area makes it particularly well suited for studies of galaxy formation across cosmic time.
We chose it for comparison with NASIM because it is one of the deepest VISTA surveys with publicly released reduced imaging.
We use the publicly available VIDEO images, which the modular CASU pipeline initially processed as part of the VDFS.
Although the reduction procedures largely follow established VISTA methods, the sky subtraction strategy is adjusted to improve sensitivity to faint sources: background frames are constructed from several unaligned, jittered paw-prints, excluding the target frame to mitigate self-subtraction, and masking only the brightest objects \citep{Jarvis13}.
This approach enhances the recovery of faint compact sources, but can lead to systematic over-subtraction around undetected or diffuse features, particularly in the LSB regime.

From the VIDEO XMM-LSS field, we select the tile \texttt{xmm2}, which includes the nearby galaxy NGC\,895, located at R.A.(J2000) = $02^{\mathrm{h}}21^{\mathrm{m}}36^{\mathrm{s}}$.47, Dec(J2000) = $-05^\circ31'17''$.02.
Although positioned near the edge of the tile, NGC\,895's large angular extent and proximity make it an ideal test case for evaluating the recovery of extended LSB features.
In Figure \ref{fig:nasim-casu}, we present a side-by-side visual comparison between the publicly released VIDEO image (top panel) and the NASIM-processed version (bottom panel).
Differences in background uniformity and the treatment of extended structures are clear.
In the VDFS reduction, the outer regions of NGC\,895 appear significantly truncated, with evident signs of over-subtraction near the galaxy’s edges.
We also detect ghost-like residuals symmetrically flanking the galaxy, likely caused by persistence effects associated with the presence of a bright, extended object.

To better illustrate these effects, we include zoom-in panels for both reductions, highlighting NGC\,895 and a nearby dwarf galaxy (LEDA 3098171).
These close-ups enable us to make a more detailed visual comparison of how each pipeline preserves extended LSB emission.

We quantitatively confirm the improvements achieved with NASIM in Figure \ref{fig:sbngc895}, where we present the surface brightness profiles of NGC\,895, extracted from both reductions.
For NGC\,895, the NASIM profile extends smoothly and significantly further into the outskirts, maintaining the galaxy’s radial structure down to fainter levels.

\begin{figure}
  \begin{center}
  \ifdefined\makepdf%
    \tikzsetnextfilename{figure-sb-dwarf}%
%
%

\begin{tikzpicture}

  \usetikzlibrary {decorations.text}

  \scriptsize

  \draw [white](-1cm,-0.8cm) rectangle (8cm,7.4cm);

    \begin{axis}[
        xlabel={R (arcsec)},
        ylabel={$\mu_{\mathrm{K_s}}$ (mag arcsec$^{-2}$)},
        legend pos=outer north east,
        width=9.3cm,
        height=8.8cm,
        ymin=22.2,
        ymax=28.9,
        y dir=reverse,
        xmin=0, xmax=34.95,
        xtick={5,10,15,20,25,30,35},
        xticklabels={5,10,15,20,25,30,35},
        grid=both,
        grid style={line width=.1pt, draw=gray!20},
        major grid style={line width=.2pt, draw=gray!20},
        legend style={
            at={(0.77,0.96)}, 
            anchor=north, 
            draw=black, 
            fill=white, 
            text height=1.5ex, 
            text depth=.25ex, 
            column sep=1ex, 
        }
    ]

    \addplot[
        color=purple,
        mark=square*,
        mark size=1.4pt,
        only marks,
        restrict y to domain=0:30,
        error bars/.cd,
        y dir=both,
        y explicit,
    ] table[x index=0, y index=2, y error index=3, col sep=space]
            {tex/build/texts/sb-k-nasim-dwarf.txt};
    \addlegendentry{NASIM Pipeline}

    \addplot[
        color=teal,
        mark=*,
        mark size=1.5pt,
        only marks,
        restrict y to domain=0:28.9,
        error bars/.cd,
        y dir=both,
        y explicit,
    ] table[x index=0, y index=2, y error index=3, col sep=space]
            {tex/build/texts/sb-k-casu-dwarf.txt};
    \addlegendentry{Standard Pipeline}

    \end{axis}


\end{tikzpicture}%
  \else
    \includegraphics[width=0.99\linewidth]{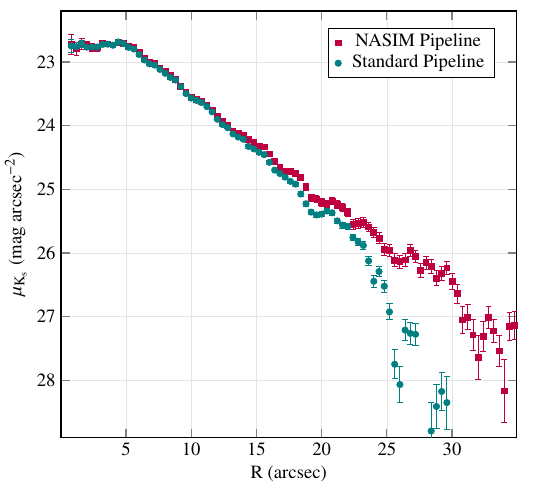}
  \fi

  \end{center}
  \vspace{-3mm}
  \caption{\label{fig:sbdwarf}
  Comparison of surface brightness profiles for the selected dwarf galaxy (LEDA 3098171), extracted from the standard VIDEO reduction and the NASIM-processed data. The NASIM reduction reveals the galaxy’s LSB outskirts, which are suppressed in the standard pipeline. This enhanced recovery of faint stellar light leads to a more complete accounting of the galaxy’s total luminosity and stellar mass.
  }
\end{figure}
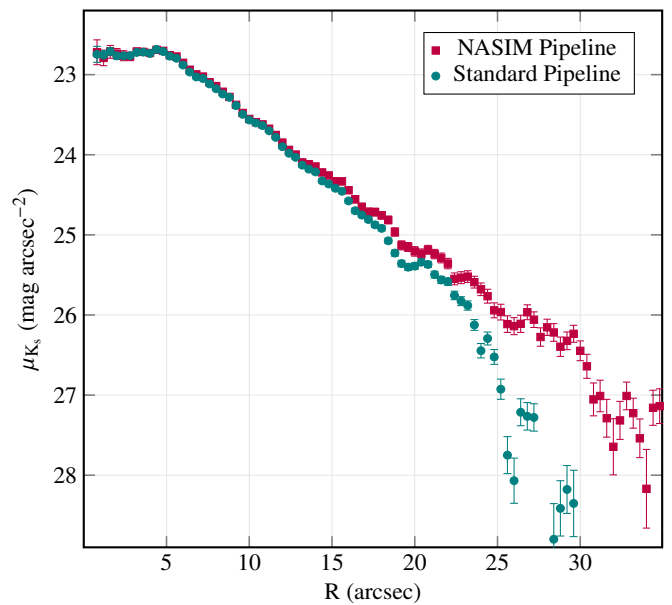

The dwarf system (Fig. \ref{fig:sbdwarf}) also exhibits a striking difference: while both reductions detect its central core, only the NASIM-processed data preserves the faint outer envelope.
By better recovering the outskirts of dwarf galaxies, NASIM enables a more accurate estimation of their total stellar mass, capturing light from extended LSB components that would otherwise be missed.

The conservative sky subtraction approach used in NASIM to preserve LSB features does not significantly compromise image depth:
the $5\sigma$ limiting magnitude in 2 arcsec diameter apertures is 23.67 mag for NASIM, compared to 23.72 for the default VIDEO reduction.
This demonstrates that NASIM maintains competitive sensitivity to faint compact sources, while simultaneously enabling the robust recovery of diffuse LSB structures.
Because the creation of source catalogues with NASIM is beyond the scope of this paper, we do not comment here on whether, and to what extent, the NASIM results can be used for the study of higher-redshift galaxies for which the VIDEO data reduction was optimised. In future work, we will explore quantitatively whether a single NASIM-inspired pipeline can be used to produce both point source- and LSB-optimised imaging results.

This comparison highlights a central strength of NASIM, which is to meet the stringent demands of LSB science while accurately recovering faint, extended features.
Our tailored design enables the reliable detection and preservation of diffuse structures such as galaxy outskirts, stellar haloes, LSB galaxies, tidal features, and ICL—domains where conventional pipelines often prove inadequate.
Having presented the unique features of our pipeline and validated them in this paper, our future work will focus on releasing the full KEDFS dataset along with associated LSB-optimised catalogues.
In addition, we plan to process and release large portions of the VHS using NASIM in the near future.
While the VHS is shallower than the KEDFS (see Table \ref{tab:vista}), its vast sky coverage makes it uniquely valuable for LSB studies at scale—offering the potential to transform our understanding of diffuse structures across the NIR sky.

The methods we implement in NASIM—such as advanced sky modelling, mitigation of detector artefacts, and precise treatment of noise features—are broadly applicable beyond VISTA.
We can adapt them to other NIR and infrared instruments, including operational and upcoming space-based missions like Euclid, Roman, and ARRAKIHS\footnote{\url{https://www.arrakihs-mission.eu}} (Analysis of Resolved Remnants of Accreted galaxies as a Key Instrument for Halo Surveys) which face similar challenges such as 1/f noise, persistence, and background instabilities.

Crucially, we design NASIM as a fully open, modular, and reproducible workflow, aligned with best practices in transparent scientific computing.
In an era increasingly shaped by large surveys and complex pipelines, we believe such open frameworks are essential for building confidence in the data and fostering collaborative progress across the astronomical community.

\section {Summary} \label{sec:summary}

In this paper, we address the challenge of studying LSB structures in ground-based NIR observations—a regime that remains relatively unexplored due to strong atmospheric absorption and telluric emission.
To tackle this, we develop NASIM: a fully automated and reproducible data reduction pipeline specifically designed to preserve diffuse emission in VISTA/VIRCAM observations.
We incorporate several key techniques implemented through Gnuastro, including a novel flat-fielding approach (adaptive flat) that effectively removes large-scale instrumental patterns, a non-aggressive sky subtraction strategy tailored for extended emission, and robust outlier rejection during weighted coaddition.

While NASIM applies to all VISTA/VIRCAM surveys, this paper focusses on KEDFS, a newly introduced deep dataset in the $K_s$ band (the $5\sigma$ limiting magnitude in 2 arcsec diameter apertures is $\sim$23.5 mag), covering approximately 20 deg$^2$ of the sky.
Our NASIM reduction of KEDFS achieves a depth of $\sim$27.7 mag arcsec$^{-2}$ ($3\sigma$ over 100 arcsec$^2$), revealing prominent LSB features.
As a validation, we explore several LSB structures in selected KEDFS tiles—including the outskirts of NGC\,1494 galaxy, an example UDG: \mbox{SMDG,J0406539--480442}, and ICL in the galaxy cluster ACT-CLJ0414.2$-$4612.
We plan a full reduction of the KEDFS dataset along with the development of associated LSB-optimised catalogues.
Our comparison with the default VISTA/VIRCAM pipeline on VIDEO survey data demonstrates that NASIM not only recovers diffuse LSB structures typically missed by other pipelines but also preserves competitive sensitivity to faint compact sources.

Looking forwards, we anticipate that NASIM will serve as a foundation for future LSB studies in the NIR, particularly in synergy with space-based observatories such as Euclid, Roman, and ARRAKIHS where legacy ground-based imaging remains an essential complement.
This work also underscores the value of fully open and transparent workflows as a path towards more reliable, reproducible science in the era of large surveys.

\section*{Code and data availability}

This project was developed in the reproducible framework of Maneage\footnote{\url{https://maneage.org}} \citep[\emph{Man}aging data lin\emph{eage},][latest Maneage commit \texttt{\maneageversion{}}, from \maneagedate]{maneage}.
The project was built on an {\machinearchitecture} machine with {\machinebyteorder} byte-order, see Appendix \ref{appendix:software} for the used software and their versions.
This paper is created from the Git commit \texttt{\projectversion}, hosted on \href{https://gitlab.com/nasim-projects/pipeline}{Gitlab} repository (\texttt{\small{\projectgitbranch}} branch); archived on Software Heritage for long-term preservation: \href{https://archive.softwareheritage.org/swh:1:dir:b3657cfb6053fd976695bd63c15cb99e5095648a;origin=https://gitlab.com/nasim-projects/pipeline;visit=swh:1:snp:ab7c6f0b9999f42d77154103c1bc082fa23b325c;anchor=swh:1:rev:afeb282c01983cba2a11eb4b2f25d5a40d35c164}{swh:1:dir:b3657cfb6053fd976695bd63c15cb99e5095648a}.
All of the files necessary to reproduce this paper — including the source, software tarballs, and essential input data — are archived on Zenodo: \href{http://doi.org/10.5281/zenodo.16152699}{16152699}.
In addition, some stacked FITS files from the KEDFS survey released by NASIM are also available on Zenodo.


\begin{acknowledgements}

The authors wish to express their heartfelt gratitude to the late Dr Mario Nonino, who led the original VISTA proposal for the KEDFS. His dedication, insight, and generosity left a lasting impact on this project and on all who had the privilege of working with him. We dedicate this paper to his memory.
ES wishes to thank Pedram Ashofteh Ardakani, Ra{\'u}l Infante-Sainz, Giulia Golini, and Zahra Sharbaf for their useful help, and suggestions in code writing.
Co-funded by the European Union (MSCA Doctoral Network EDUCADO, GA 101119830 and Widening Participation, ExGal-Twin and UNDARK, GAs 101158446 and 101159929). JHK and IT acknowledges grants PID2022-136505NB-I00 and PID2022-140869NB-I00, respectively, funded by MCIN/AEI/10.13039/501100011033 and EU, ERDF.
HD acknowledge support from the Agencia Estatal de Investigación del Ministerio de Ciencia, Innovación y Universidades (MCIU/AEI) under grant (Construcción de cúmulos de galaxias en formación a través de la formación estelar oscurecida por el polvo) and ERDF with reference (PID2022-143243NB-I00/10.13039/501100011033).
SC, and MA acknowledge support from the AEI-MICIU of the Spanish Ministry of Science, Innovation, and Universities under grants PID2023-149139NB-I00, and PID2021-124918NA-C43, respectively, co-funded by the ERDF where applicable.

\end{acknowledgements}

\bibliographystyle{aa}
\bibliography{references}

\begin{appendix}

\section{Astrometric calibration details} \label{app:astrometry}

Astrometry in NASIM is conducted in two stages.
Initially, a rapid astrometric solution is achieved using the \textsc{Astrometry.net} program \citep{Lang10}.
The steps are as follows:

\begin{enumerate}

\item
  Download \textsc{Gaia} DR3 \citep{Gaia21} as a reference catalogue.
  To streamline handling, the sky is partitioned into smaller files.
  By using an initial estimate of image locations, \textsc{Astrometry.net} optimises the search by skipping non-overlapping sky tiles, resulting in faster and more memory-efficient processing.
We encoded the coordinates and radius of the EDFS region into a configuration file.

\item
  Build index files from the reference catalogue to expedite the matching process and enable parallel execution.

\item
  Generate catalogues from images as input for \textsc{Astrometry.net}.
  By default, \textsc{Astrometry.net} uses an internal catalogue with predefined parameters.
  However, we opted to generate external catalogues, incorporating settings that minimise the occurrence of unreliable and spurious detections.
  To create these catalogues, we employed Gnuastro's \textsc{NoiseChisel} and \textsc{Segment} programs to generate clump maps of sources.
  Subsequently, the \textsc{MakeCatalog} program was used to measure the brightness of these clumps \citep{Akhlaghi19b}.

\item
  Solve the astrometry of images using the generated index files.
  \textsc{Astrometry.net} provides a non-linear mapping from image to sky coordinates through the SIP (Simple Imaging Polynomial) projection.
  We then used Gnuastro's \textsc{Fits} and \textsc{Arithmetic} programs to convert the SIP solution into a TPV (TAN projection with polynomial distortion correction) convention, and transferred the WCS headers into the original, non-astrometrised images.

\end{enumerate}

\begin{figure}
  \begin{center}
  \ifdefined\makepdf%
    \tikzsetnextfilename{figure-astrometry}%
    \input{tex/src/figure-astrometry.tex}%
  \else
    \includegraphics[width=0.95\linewidth]{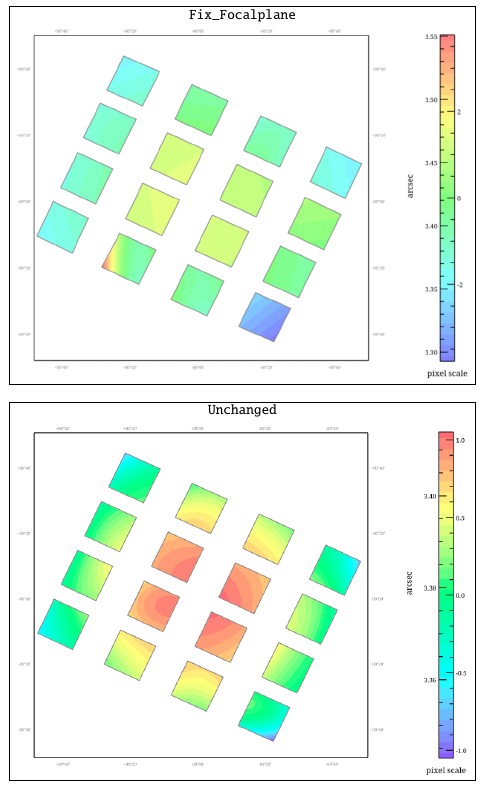}
  \fi

  \end{center}
  \vspace{-3mm}
  \caption{\label{fig:astrometry}
    Distortion maps of the 16 VIRCAM detectors after astrometric calibration with \textsc{SCAMP}. Each panel is labelled with the calibration setting used: \texttt{Fix\_Focalplane} (top) and \texttt{Unchanged} (bottom). The distortion pattern in the \texttt{Unchanged} case is consistent with the expected optical behaviour of a wide-field telescope, with minimal distortion near the centre and increasing distortion towards the edges. Pixel scale variations are colour-coded, with the corresponding scale bars shown to the right of each panel.
  }
\end{figure}

Using Gnuastro's \textsc{astscript-ds9-region} program and the \textsc{Gaia} catalogue, we verified the accuracy of WCS coordinates for numerous images.

Following the initial rapid astrometric solution with \textsc{Astrometry.net}, a more refined calibration was necessary to correct for small-scale distortions and ensure precise alignment across all detectors.
This refinement was achieved through a second stage of astrometry using the \textsc{SCAMP} program \citep{Bertin06}, which builds upon the initial solution by modelling and correcting residual distortions based on source catalogues.
\textsc{SCAMP} calculates astrometric projection parameters using source catalogues generated by the \textsc{SExtractor} program \citep{Bertin96}.
For catalogue creation with \textsc{SExtractor}, we utilised a multi-extension image containing all 16 detectors of VIRCAM to enhance astrometric precision.
To ensure a catalogue populated with reliable sources, we adjusted key input parameters in the \texttt{sex.config} file, such as \texttt{detect\_minarea} (the minimum number of pixels above threshold) and \texttt{detect\_thresh}, focusing on detecting only bright sources.

To achieve an accurate solution, we carefully scrutinised various parameters in the \textsc{SCAMP} configuration file.
For instance, we used a ``focal-plane'' model for VIRCAM, available within \textsc{SCAMP}, as an input header.
We initially set the \texttt{mosaic\_type} option to \texttt{fix\_focalplane} for pattern matching.
The resulting distortion map, illustrating the astrometric model of the input frames after calibration with this configuration, is shown at the top of Figure \ref{fig:astrometry} (\texttt{Fix\_Focalplane}).
It reveals limitations in aligning all detectors consistently, indicating that this configuration was not optimal for our dataset.

In pursuit of maintaining better alignment across all detectors, we determined that setting the \texttt{mosaic\_type} option to \texttt{unchanged} yielded the most suitable solution.
The improved distortion map is presented at the bottom of Figure \ref{fig:astrometry} (\texttt{Unchanged}).
The pixel scale is depicted across the field of view using a colour gradient from blue to red.
The observed distortion pattern aligns with the expected behaviour for an astronomical telescope, with minimal distortion at the centre and larger distortions towards the edges.

The computed solution from \textsc{SCAMP} follows the WCS standard and was saved as a header file.
We incorporated these headers into each image using Gnuastro’s \textsc{Fits} program.

\begin{figure*}
  \begin{center}
  \ifdefined\makepdf%
    \tikzsetnextfilename{figure-sky-std}%
    \input{tex/src/figure-sky-std.tex}%
  \else
    \includegraphics[width=0.99\linewidth]{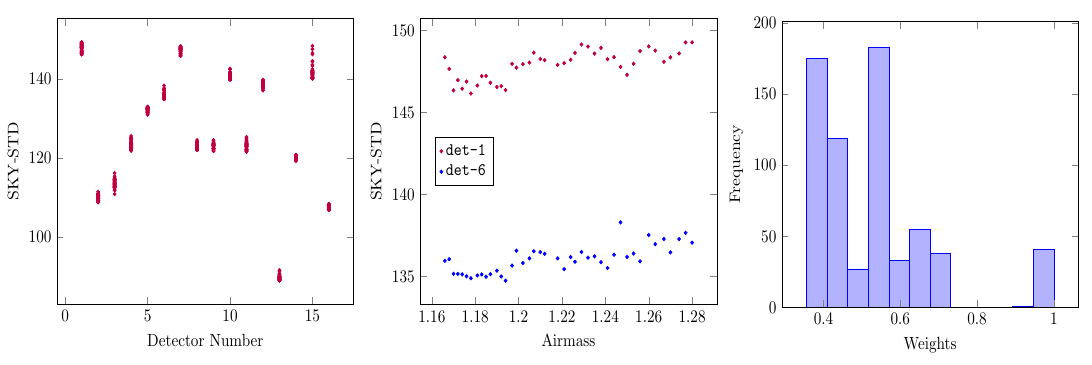}
  \fi

  \end{center}
  \vspace{-3mm}
  \caption{\label{fig:sky-std}
    Example of variations in SKY-STD across all images taken during a single night. Left panel: Variations in SKY-STD as a function of detector number, highlighting discernible differences in image quality among detectors.
    Middle panel: Correlation between SKY-STD and airmass for selected detectors, illustrating the impact of zenith angle on image quality. Right panel: Distribution of computed weights for all images from the night, reflecting the range of observational quality.
  }
\end{figure*}

\section{Image weighting details} \label{app:weighting}

Image weighting is critical to account for variations in image quality arising from changes in airmass during the night and detector-to-detector differences.
Image weights were computed by measuring the standard deviation of the sky (SKY-STD) in each frame, employing the \textsc{NoiseChisel} program without interpolation, following a similar approach to that described in Section \ref{sec:sky}.
A $\sigma$-clipped median ($3\sigma$ clipping with a tolerance of 0.1) was applied to the tiles in the SKY-STD frames produced by \textsc{NoiseChisel}.
Weights were normalised such that the highest-quality data (with the minimum SKY-STD) were assigned a weight of 1, while the lowest-quality data (with the maximum SKY-STD) were assigned a weight given by

\begin{equation}
\text{weight} = \frac{(\text{SKY-STD}_{\text{min}})^2}{(\text{SKY-STD}_{\text{max}})^2}\,.
\end{equation}

An example of the variation in SKY-STD across a single night is shown in Figure \ref{fig:sky-std}.
The left panel illustrates fluctuations in SKY-STD as a function of detector number, revealing notable quality differences among the detectors.
The middle panel shows the variation of SKY-STD with airmass for selected detectors, demonstrating a clear correlation with atmospheric conditions.
The right panel presents the resulting distribution of weights, which range from 0.67 to 1.0, reflecting the variability in observation quality throughout the night.

\begin{figure*}
  \begin{center}
  \ifdefined\makepdf%
    \tikzsetnextfilename{figure-zp-diffmag}%
    \input{tex/src/figure-zp-diffmag.tex}%
  \else
    \includegraphics[width=0.85\linewidth]{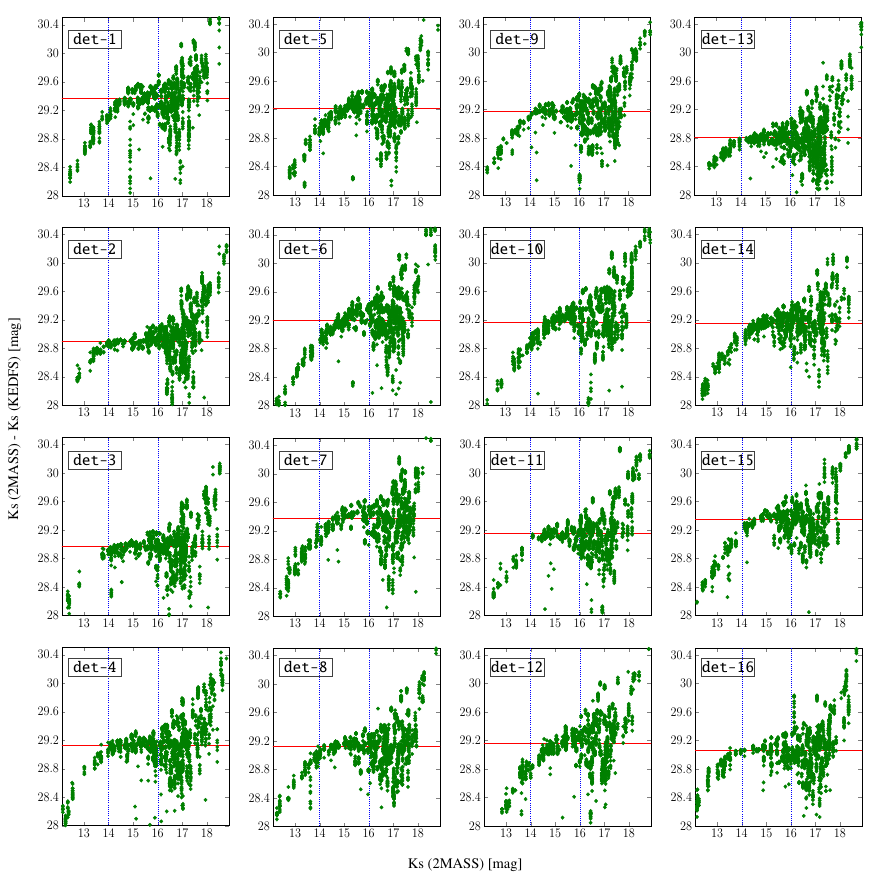}
  \fi

  \end{center}
  \vspace{-3mm}
  \caption{\label{fig:zpdiffmag}
    Difference between 2MASS and KEDFS magnitudes as a function of 2MASS magnitude, measured within a 2 arcsec radius aperture, for 42 frames taken over a single night across all detectors. The photometric calibration is based on sources with 2MASS AB magnitudes between 14 and 16.
    }
\end{figure*}

\section{Photometric calibration details} \label{app:photometry}

\begin{figure*}
  \begin{center}
  \ifdefined\makepdf%
    \tikzsetnextfilename{figure-zp-histo}%
    \input{tex/src/figure-zp-histo.tex}%
  \else
    \includegraphics[width=0.85\linewidth]{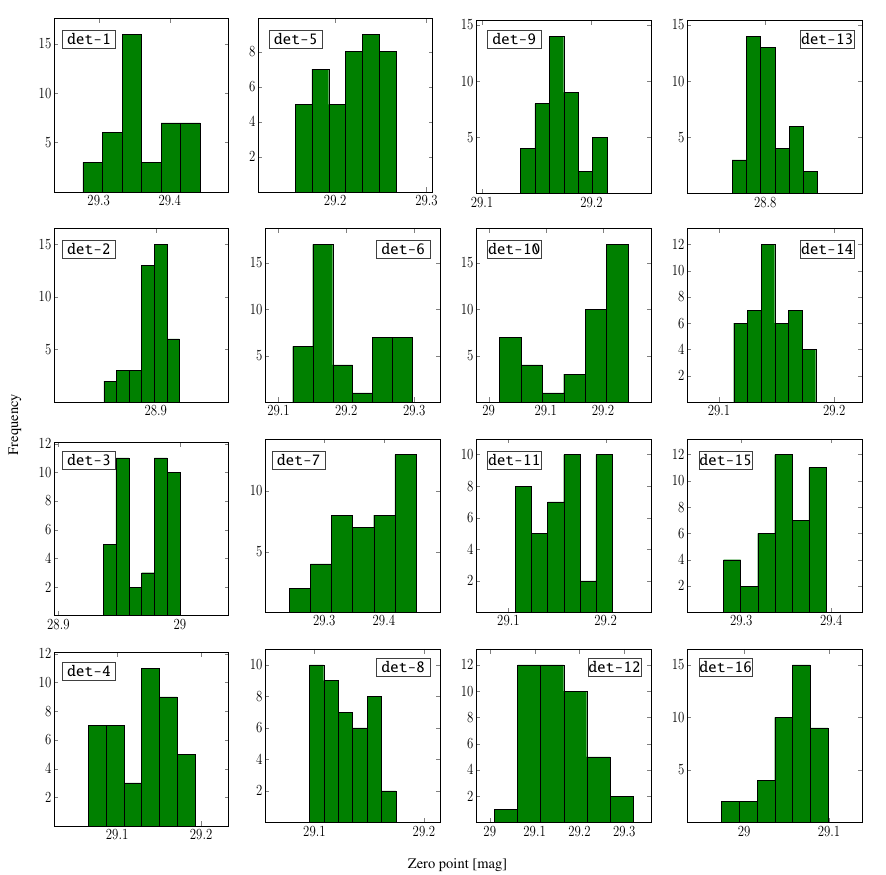}
  \fi

  \end{center}
  \vspace{-3mm}
  \caption{\label{fig:zphisto}
    Histogram of photometric zero points derived for 42 images obtained over a single night, covering all 16 detectors. Zero points were estimated using $\sigma$-clipping ($3\sigma$ with a tolerance of 0.1), based on calibration sources with AB magnitudes between 14 and 16 from the 2MASS catalogue.
   }
\end{figure*}

The 2MASS catalogue was adopted as the reference for the photometric calibration of our images.
To enable direct comparison, 2MASS magnitudes were converted from the Vega to the AB system \citep{Blanton07}.
Aperture photometry on the processed images was performed using Gnuastro’s \textsc{MakeProfile} and \textsc{MakeCatalog} programs, with stellar positions accurately determined relative to the Gaia catalogue.
The KEDFS catalogues were then matched to the 2MASS reference using Gnuastro’s \textsc{Match} program. The method used for estimating the photometric zero point follows the same approach developed for Gnuastro’s \textsc{astscript-zeropoint} program \citep{Eskandarlou23}.

Figure \ref{fig:zpdiffmag} shows the difference between 2MASS and KEDFS magnitudes, plotted against the 2MASS magnitudes, measured within a 2-arcsec radius aperture across 42 images from a single night and for different detectors.
Due to the completeness limits of the 2MASS catalogue and the saturation levels in VISTA images, the calibration was restricted to sources with magnitudes between 14 and 16 mag (AB system) in 2MASS.
To determine the optimal aperture size for photometry, we initially tested various aperture radii ranging from 1 to several times the FWHM.
For each case, we computed the standard deviation of the derived zero points and selected the aperture size that minimised this dispersion, corresponding to a radius of 2 arcsec.
This choice was found to be consistent across all detectors.

The distribution of zero points obtained from the 42 images across all 16 detectors is shown in Figure \ref{fig:zphisto}.
Zero points were calculated using Gnuastro's $\sigma$-clipping algorithm, adopting a $3\sigma$ threshold with a tolerance of 0.1.
To assess the robustness of the photometric calibration, we examined the correlation between the derived zero points and the airmass at the time of observation.
The typical variation of zero points across the range of airmasses observed is approximately 0.1 mag.
A weak but statistically significant trend was observed, with zero points decreasing as airmass increased, indicative of increased atmospheric extinction at higher airmasses.
Overall, the scatter in the zero point distribution is consistent with expectations, primarily reflecting variations in observing conditions.
Finally, the photometric zeropoints of all images were normalised to a constant value of 27 mag.

\section{Resampling details} \label{app:resampling}

Image resampling was performed using Gnuastro's \textsc{Warp} program.
\textsc{Warp} employs a non-parametric interpolation technique known as ``pixel mixing,'' which uses the fractional area of an output pixel over an input pixel to estimate how much each input pixel should contribute to an output pixel.
By default, \textsc{Warp} aligns the pixel grid of the input image with the WCS coordinates embedded in the header.
Given that the pixel scale of VIRCAM detectors varies both between detectors and across individual frames, we employed the \texttt{cdelt} option to standardise the pixel size to 0.34\,arcsec.
After testing different pixel scales, we adopted a final value of 0.2\,arcsec, as it provides finer sampling of the PSF and facilitates more accurate measurements of LSB structures\footnote{\url{https://www.gnu.org/software/gnuastro/manual/html_node/Moire-pattern-and-its-correction.html}}.
This choice follows the convention adopted in several other VISTA surveys, such as the VIDEO survey.
Nevertheless, the pipeline remains flexible, and the previously tested value of 0.34\,arcsec can still be used where computational resources or storage limitations are more critical.

Due to the large size of a VISTA tile (exceeding 1.5 square degrees; \citealt{Cross12}), we opted not to resample the entire tile.
Instead, we selected a subregion of approximately 0.3 square degrees centred on the target coordinates.
This choice was motivated by limitations in available RAM, as well as the computational challenges associated with handling large image mosaics.

Following resampling, the corresponding weight values for each image (Section \ref{sec:weight}) were applied.
Since these weights are scalar values, unaffected by rotation, warping, or sky curvature, they were simply multiplied after the warping step.
This approach significantly reduced storage, memory usage, and processing time; otherwise, it would have been necessary to warp both the science images and their associated weight maps.
Examples of a resampled science image (\texttt{Resampled image}) and its corresponding weight image (\texttt{Weight image}) are shown in Figure \ref{fig:flowchart}, although the latter is generated at the stacking stage by the \texttt{stack.mk} Makefile after outlier rejection (see Section \ref{sec:stack}).

\section{Software acknowledgement}
\label{appendix:software}

This research was done with the following free software programs and libraries:  1.23, Astrometry.net 0.91 \citep{astrometrynet}, Boost 1.77.0, Bzip2 1.0.8, Cairo 1.16.0, cdsclient 3.84, CFITSIO 4.1.0, CMake 3.24.0, cURL 7.84.0, Dash 0.5.11-057cd65, Discoteq flock 0.4.0, Eigen 3.4.0, Expat 2.4.1, FFTW 3.3.10 \citep{fftw}, File 5.42, Fontconfig 2.14.0, FreeType 2.11.0, Git 2.37.1, GNU Astronomy Utilities 0.23 \citep{gnuastro}, GNU Autoconf 2.71, GNU Automake 1.16.5, GNU AWK 5.1.1, GNU Bash 5.2-rc2, GNU Binutils 2.39, GNU Bison 3.8.2, GNU Compiler Collection (GCC) 12.1.0, GNU Coreutils 9.1, GNU Diffutils 3.8, GNU Emacs 28.1, GNU Findutils 4.9.0, GNU gettext 0.21, GNU gperf 3.1, GNU Grep 3.7, GNU Gzip 1.12, GNU Integer Set Library 0.24, GNU libiconv 1.17, GNU Libtool 2.4.7, GNU libunistring 1.0, GNU M4 1.4.19, GNU Make 4.3, GNU Multiple Precision Arithmetic Library 6.2.1, GNU Multiple Precision Complex library, GNU Multiple Precision Floating-Point Reliably 4.1.0, GNU Nano 6.4, GNU NCURSES 6.3, GNU Readline 8.2-rc2, GNU Scientific Library 2.7, GNU Sed 4.8, GNU Tar 1.34, GNU Texinfo 6.8, GNU Wget 1.21.2, GNU Which 2.21, GPL Ghostscript 9.56.1, Help2man , Less 590, Libffi 3.4.2, Libgit2 1.3.0, libICE 1.0.10, Libidn 1.38, Libjpeg 9e, Libpaper 1.1.28, Libpng 1.6.37, libpthread-stubs (Xorg) 0.4, libSM 1.2.3, Libtiff 4.4.0, libXau (Xorg) 1.0.9, libxcb (Xorg) 1.15, libXdmcp (Xorg) 1.1.3, libXext 1.3.4, Libxml2 2.9.12, libXt 1.2.1, LibYAML 0.2.5, Lzip 1.23, OpenBLAS 0.3.24, Open MPI 4.1.1, OpenSSL 3.0.5, PatchELF 0.13, Perl 5.36.0, Pixman 0.40.0, pkg-config 0.29.2, PLplot 8.44, podlators 4.14, Python 3.10.6, SCAMP 2.10.0 \citep{scamp}, SExtractor 2.25.0 \citep{sextractor}, SWarp 2.41.5 \citep{swarp}, Swig 4.0.2, util-Linux 2.38.1, util-macros (Xorg) 1.19.3, WCSLIB 7.11, X11 library 1.8, XCB-proto (Xorg) 1.15, xorgproto 2022.1, xtrans (Xorg) 1.4.0, XZ Utils 5.2.5 and Zlib 1.2.11.
Within Python, the following modules were used: Astropy 5.1 \citep{astropy2013,astropy2018},  BeautifulSoup 4.10.0, Cython 0.29.24 \citep{cython2011},  Extension-Helpers 0.1,  HTML5lib 1.0.1,  Jinja2 3.0.3,  MarkupSafe 2.0.1, Numpy 1.21.3 \citep{numpy2020},  Packaging 21.3,  pybind11 2.5.0,  PyERFA 2.0.0.1,  PyParsing 3.0.4,  PyYAML 5.1, Scipy 1.7.3 \citep{scipy2020},  Setuptools 58.3.0,  Setuptools-scm 3.3.3,  Six 1.16.0,  SoupSieve 1.8 and  Webencodings 0.5.1.
The \LaTeX{} source of the paper was compiled to make the PDF using the following packages: biber 2.20, biblatex 3.20, bitset 1.3, caption 68425 (revision), courier 61719 (revision), csquotes 5.2o, datetime 2.60, ec 1.0, etoolbox 2.5l, fancyhdr 5.2, fancyvrb 4.5c, fmtcount 3.10, fontaxes 2.0.1, footmisc 7.0b, fp 2.1d, helvetic 61719 (revision), kastrup 15878 (revision), letltxmacro 1.6, lineno 5.5, logreq 1.0, mweights 53520 (revision), newtx 1.756, pdfescape 1.15, pdftexcmds 0.33, pgf 3.1.10, pgfplots 1.18.1, preprint 2011, setspace 6.7b, sttools 3.1, tex 3.141592653, texgyre 2.501, times 61719 (revision), titlesec 2.17, trimspaces 1.1, txfonts 15878 (revision), ulem 53365 (revision), xcolor 3.02, xkeyval 2.9, xpatch 0.3 and xstring 1.86.
We are very grateful to all their creators for freely  providing this necessary infrastructure. This research  (and many other projects) would not be possible without them.

\end{appendix}
\end{document}